\documentclass[%
 reprint,
 amsmath,amssymb,
 aps,
 onecolumn,
 letterpaper,
 11pt,
]{revtex4-2}

\usepackage{graphicx}
\usepackage{dcolumn}
\usepackage{bm}
\usepackage[colorlinks=true,linkcolor=blue,filecolor=blue,urlcolor=blue,citecolor=blue,pdftex,plainpages=false]{hyperref}
\usepackage{mathtools}
\usepackage[percent]{overpic}
\usepackage{booktabs}
\usepackage[rightcaption]{sidecap}
\sidecaptionvpos{figure}{t}
\usepackage{IEEEtrantools}
\usepackage{ulem}
\usepackage{cancel}
\usepackage{xcolor}

\begin{document}


\title{Highly improved staggered quarks on anisotropic lattices}

\author{Alexei Bazavov}
 \affiliation{
Department of Computational Mathematics, Science and Engineering,
Michigan State University, East Lansing, MI 48824, USA
}
 \affiliation{
Department of Physics and Astronomy,
Michigan State University, East Lansing, MI 48824, USA
}

\author{Yannis Trimis}%
 \affiliation{
	Department of Physics and Astronomy,
	Michigan State University, East Lansing, MI 48824, USA
}

\author{Johannes Heinrich Weber}
\affiliation{
Institut f\"{u}r Kernphysik, Technische Universit\"{a}t Darmstadt, Schlossgartenstraße 2, 64289 Darmstadt, Germany}
\affiliation{
Werner-Heisenberg-Gymnasium Bad D\"{u}rkheim, Kanalstraße 19, 67098 Bad D\"{u}rkheim, Germany
}

\date{\today}

\begin{abstract}
We present a study of tuning of the anisotropic highly improved staggered quark (aHISQ) action on pure gauge ensembles with the renormalized anisotropy ranging from 1 to 8. We discuss multiple gradient flow schemes for tuning the gauge anisotropy and comment on what scheme may be optimal for anisotropic simulations. Next, we compare tuning of the fermion anisotropy for the naive staggered and aHISQ actions. Finally, we study the dependence of the staggered pion taste mass splittings on anisotropy for the two actions and develop an empirical model that captures the main features of the aHISQ spectrum. We observe qualitatively different behavior of the naive and aHISQ taste spectrum with anisotropy.
\end{abstract}

\maketitle


\section{Introduction}
\label{sec_intro}
The main motivation for the present work is to develop a formalism to generate gauge field ensembles with optimal characteristics for spectral reconstruction. One of the physics goals is to study the melting pattern of heavy quarkonia and heavy open-flavor states in quark-gluon plasma which are fully encoded in the temperature dependence of their spectral functions.
Reconstruction of the spectral functions from the Euclidean correlation functions computed in lattice QCD is an ill-posed problem which, despite of the two decades~\cite{Asakawa:2000tr} of effort, remains challenging.
Recent work on spectral reconstruction at finite temperature can be found in Refs.~\cite{Aarts:2014cda,Kim:2014iga,Ding:2017std,Ilgenfritz:2017kkp,Ding:2018uhl,Kim:2018yhk,Chen:2021giw,Huang:2023gpb,Ali:2024xae,Bignell:2025ddh,Skullerud:2025iqt,Takahashi:2025ujh,Lombardo:2025sfo,Skullerud:2026sek}, and
while some progress has been made with the reconstruction methods
(recent attempts include reconstruction with the Maximum Entropy Method~\cite{Bryan1990,JARRELL1996133}, Backus-Gilbert method~\cite{Backus:1968} and its modifications~\cite{Hansen:2019idp}, a recently formulated Bayesian Reconstruction method~\cite{Burnier:2013nla}, and stochastic-based analytic continuation techniques~\cite{PhysRevB.62.6317,PhysRevE.81.056701}), it is of primary importance to produce gauge field ensembles that are specifically tuned for addressing the reconstruction problem.
Namely, one is looking for ensembles with (potentially) very high statistics and large temporal extents $N_\tau$. Isotropic ensembles with large $N_\tau$ are prohibitively expensive, however, making lattices anisotropic, \textit{i.e.}, $a_{\sigma}/a_{\tau}=\xi\sim 6 - 8$, where $a_{\sigma}$ ($a_{\tau}$) is the spatial (temporal) lattice spacing, to large degree alleviates that problem.
In terms of statistics, staggered fermions have proven to be the leading formulation for finite-temperature QCD with some ensembles of order a hundred thousand configurations.
Naive (\textit{i.e.}, unimproved) staggered fermions feature large discretization effects in the hadron spectrum, and early attempts with anisotropic naive staggered fermions~\cite{Nomura:2004qu, Levkova:2006gn} have not been further pursued.
However, development of improved actions such as asqtad~\cite{Blum:1996uf,Orginos:1999cr}, p4~\cite{Heller:1999xz}, stout~\cite{Morningstar:2003gk} and HISQ~\cite{Follana:2006rc} allowed one to significantly suppress the discretization (primarily, taste-breaking) effects in the fermion sector.
We therefore explore an anisotropic variant of the HISQ action that we call anisotropic Highly Improved Staggered Quarks (aHISQ).

It is important to note that anisotropic ensembles with dynamical fermions are used for the spectral function reconstruction problem by the FASTSUM collaboration, \textit{e.g.}, Refs.~\cite{Aarts:2014cda,Lombardo:2025sfo,Skullerud:2025iqt,Skullerud:2026sek}. However, they are based on the fixed scale approach with Wilson fermions. Wilson fermions are more computationally demanding and therefore are inherently more limited in statistics. The fixed scale approach, while being more economical for tuning, leads to decreasing the temporal extent $N_\tau$ with increasing temperature. 
In the temperature range relevant for the physics of quark-gluon plasma those ensembles at present have $N_\tau\le 36$. Such temporal extents are in the range that can be reached today on isotropic lattices with staggered fermions, \textit{e.g.}, Ref.~\cite{Bazavov:2023dci}, and thus one should be able to reach much larger $N_\tau$ with an anisotropic staggered fermion formulation.

The paper is organized as follows. In Sec.~\ref{sec_QCD_ani} we present the methodology of tuning the parameters of the gauge and fermion action, and also discuss deformation of the staggered meson spectrum when anisotropy is introduced.
In Sec.~\ref{sec_dyn_ani} we provide a full algorithm for the aHISQ fermion force, necessary for implementation of the anisotropic Rational Hybrid Monte Carlo updating algorithm. Sec.~\ref{sec_num_res} constitutes the main part of the paper where all numerical results are presented which include tuning of the gauge and fermion actions for the renormalized anisotropy ranging from 1 to 8, measurements of the pion taste mass splittings, and development of a model that describes the main observed features of the aHISQ taste spectrum and its dependence on anisotropy. The conclusions are presented in Sec.~\ref{sec_concl} and the appendices document the numerical data on the pure gauge ensembles, tuning of the aHISQ action, aHISQ pion taste masses and the validation tests against existing results in the literature.

\section{QCD on anisotropic lattices}
\label{sec_QCD_ani}
Anisotropic lattices have been used for studies in lattice QCD since the early days~\cite{HASENFRATZ1981210,Karsch:1982ve}. The anisotropy is introduced naturally within the path-integral formulation of a quantum field theory on a discretized spacetime grid, as the temporal and spatial lattice spacings appear for different reasons, and therefore need not be equal, \textit{e.g.}, Ref.~\cite{Rothe:1992nt}.
The introduction of anisotropy adds complexity to the tuning of lattice ensembles, since the tuning of the couplings and quark masses is now entangled with the anisotropy parameters that appear in the action. For that reason, anisotropic simulations are less common than the isotropic ones. The recent work employing anisotropic ensembles includes thermodynamics, glueball and hadronic spectroscopy and spectral reconstruction~\cite{Smecca:2024gpu,Jiang:2022ffl,Lombardo:2025sfo,Skullerud:2025iqt}. Apart from the early attempts with the anisotropic naive staggered fermions in the early 2000s~\cite{Nomura:2004qu,Levkova:2006gn}, the present large-scale anisotropic simulations employ only Wilson fermions.

\subsection{Gauge anisotropy}
\label{sec_gauge_ani}

In the following the lattice spacing in the spatial directions is denoted $a_\sigma$ and in the temporal one $a_\tau$. We refer to the ratio $\xi=a_\sigma/a_\tau$ of these dimensionful quantities as the renormalized anisotropy. To achieve different lattice spacings, one needs an additional bare parameter in the action, namely the bare gauge anisotropy, denoted here by $\xi_0$. The introduction of anisotropy in a lattice gauge action is quite standard with the details fully worked out in the early work by Karsch~\cite{Karsch:1982ve}.
We use the anisotropic tree-level Symanzik-improved gauge action 
\begin{eqnarray}\label{eq_gauge_action}
S_g&=&\beta\frac{1}{\xi_0}
\left(
c_P\sum_{P_{\sigma\sigma}}
\left(1-\frac{1}{3}{\rm Re}{\rm Tr}(P_{\sigma\sigma})\right)
+
c_R\sum_{R_{\sigma\sigma}}
\left(1-\frac{1}{3}{\rm Re}{\rm Tr}(R_{\sigma\sigma})\right)
\right)
\nonumber\\
&+&
\beta\xi_0\left(
c_P\sum_{P_{\sigma\tau}}
\left(1-\frac{1}{3}{\rm Re}{\rm Tr}(P_{\sigma\tau})\right)
+
c_R\sum_{R_{\sigma\tau}}
\left(1-\frac{1}{3}{\rm Re}{\rm Tr}(R_{\sigma\tau})\right)
\right),
\end{eqnarray}
where the sums are over all $1\times1$ and $1\times2$ Wilson loops in all possible orientations, denoted $P$ and $R$, respectively. The indices $\sigma\sigma$ ($\sigma\tau$) represent spatial-spatial (spatial-temporal) orientations, and $c_P=1$, $c_R=-1/20$ and $\beta=10/g^2$. (This is the standard L\"{u}scher-Weisz action~\cite{Weisz:1982zw,Luscher:1984xn} $c_0=5/3$, $c_1=-1/12$ with $c_0$ combined with $\beta$ for convenience.)

Mapping the bare parameters $(\beta,\xi_0)$ to the dimensionful quantities $(a_\sigma,a_\tau)$, or, equivalently, to $(a_\sigma,\xi)$, requires an analog of the usual lattice scale setting procedure, modified to provide two quantities. In general, one computes a physical quantity that allows for separating spatial and temporal direction (\textit{e.g.}, differently oriented Wilson loops as in the Klassen method~\cite{Klassen:1998ua}) and imposes that it is the same in physical units in both directions, thus fixing $(a_\sigma,a_\tau)$. As the use of the gradient flow is now a more common and more convenient practice to set the scale~\cite{Luscher:2010iy} and anisotropy, as was first proposed in Ref.~\cite{Borsanyi:2012zr}, we also use it for determining the spatial lattice spacing and the renormalized anisotropy, as described next.

\subsection{Anisotropic gradient flow}
\label{sec_ani_flow}

The gradient flow, first introduced in the lattice context in Ref.~\cite{Luscher:2009eq} and as a scale setting procedure in Ref.~\cite{Luscher:2010iy}, was extended to anisotropic case in Ref.~\cite{Borsanyi:2012zr}. The original procedure of Ref.~\cite{Borsanyi:2012zr} was later slightly modified~\cite{Borsanyi:2018srz} to make it more computationally efficient. (The procedure of Ref.~\cite{Borsanyi:2012zr} requires several gradient flows per lattice, while the one of Ref.~\cite{Borsanyi:2018srz} only one flow.) We experimented with both procedures, and having observed no qualitative differences, preferred the computationally cheaper one of Ref.~\cite{Borsanyi:2018srz}.

Let the flow anisotropy be denoted $\xi_0^{gf}$ and the gradient flow equation for the flowed lattice gauge links $U_\mu(x,t)$ be written in the following way:
\begin{eqnarray}\label{eq_gradient_flow}
	\frac{dU_\mu}{d t}= Z_\mu U_\mu,\,\,\,\,\,\,\,
	Z_\mu(x,t)=-\sum_{\nu\neq\mu} \rho_{\mu\nu}\mathcal{P}_A\Big[U_\mu(x,t)S^\dagger_{\mu\nu}[U](x,t)\Big]
\end{eqnarray}
where $t$ is the flow time, $S_{\mu\nu}$ is the staple arising from taking the group derivative of the flow action, $\mathcal{P}_A$ denotes traceless antihermitean projection
\begin{equation}\label{eq_antihermtrcless_proj}
	\mathcal{P}_A(V)=\frac{1}{2}(V-V^\dagger)-\frac{1}{6}{\rm tr}(V-V^\dagger)
\end{equation}
and the weight factors $\rho_{\mu\nu}$
\begin{equation}
\label{eq_ani_factors}
\rho_{i4}=(\xi_0^{gf})^2,\,\,\,\,\,\rho_{ij}=\rho_{4i}=1
\end{equation}
make the flow anisotropic in the spatial and temporal directions in lattice units.

As an observable, one can choose the flowed spatial and temporal action densities, similar to the original isotropic proposal~\cite{Luscher:2010iy}:
\begin{equation}\label{eq_action_densities}
	S_{\sigma\sigma}(t)=\frac{1}{4}\sum_{x,i\ne j} F_{ij}^2(x,t)\,,\,\,\,\,\,\,\,\,\,\,
	S_{\sigma\tau}(t)=\frac{1}{2}\sum_{x,i}F_{i4}^2(x,t)
\end{equation}
with a suitable discretization of the field strength tensor, or their derivatives with respect to the flow time, since those are less affected by the discretization effects, as was first noted in Ref.~\cite{BMW:2012hcm}. In the anisotropic case the scale setting conditions for the two scales $w_{0,\sigma}$, $w_{0,\tau}$ now are
\begin{eqnarray}
	\left[t\frac{d}{dt} t^2 \langle S_{\sigma\sigma}(t) \rangle\right]_{t=w_{0,\sigma}^2} = 0.15,\,\,\,\,\,\,\,\,\,
	(\xi_0^{gf})^2\left[t\frac{d}{dt} t^2 \langle S_{\sigma\tau}(t) \rangle\right]_{t=w_{0,\tau}^2} = 0.15\,.
	\label{eq_tune1}
\end{eqnarray}
If one fixes the target renormalized anisotropy $\xi$ at some value, \textit{e.g.}, $\xi=2$, and sets the gradient flow anisotropy to that value $\xi_0^{gf}=\xi$, the only remaining free parameter is the bare gauge anisotropy $\xi_0$ (we consider the bare gauge coupling $\beta$ fixed at this stage). Then $\xi_0$ is adjusted until the following additional scale setting condition is met
\begin{equation}
\frac{w_{0,\sigma}}{w_{0,\tau}}=1.
\label{eq_tune2}
\end{equation}
When the condition~(\ref{eq_tune2}) is met, the renormalized gauge anisotropy of the ensemble generated at the bare parameters $(\beta,\xi_0)$ is defined to be that fixed value of $\xi$.
While adjusting $\xi_0$ means generating gauge field ensembles at various values of $\xi_0$ (which is needed at the tuning stage when the parameter space is not yet mapped out anyway), there is only one flow per ensemble to be run, -- at the target anisotropy $\xi$.

\subsection{Fermion anisotropy}
\label{sec_ferm_ani}

To introduce anisotropy in the fermion part of the action, we can start with a generic form of the Dirac operator. The action is
\begin{equation}
	S_f=a_\sigma^3 a_\tau\sum_{x,y}\bar{{\tilde\psi}}_x\tilde M_{xy}\tilde\psi_y
\end{equation}
where the summation is over the lattice volume and $\tilde M_{xy}=\tilde D_{xy}+m\delta_{xy}$. The color and spinor indices are suppressed for brevity. The tilde $\tilde{}$ sign indicates that the quantities are dimensionful. The Dirac operator is decomposed into the spatial and temporal parts
$\tilde D_{xy}=\tilde D_{xy,\sigma}+\tilde D_{xy,\tau}$, where all matrices are of the same dimension and in the spatial (temporal) part the temporal (spatial) gauge links are set to zero. To make the quantities dimensionless, appropriate factors of the spatial and temporal lattice spacing are introduced:
\begin{equation}
\psi=a_\sigma\sqrt{a_\tau}\,\tilde\psi,\,\,\,\,\,\,\,\,
D_{xy,\sigma}=a_\sigma\tilde D_{xy,\sigma},\,\,\,\,\,\,\,\,
\xi D_{xy,\tau}=\xi a_\tau\tilde D_{xy,\tau}=a_\sigma\tilde D_{xy,\tau}.
\end{equation}
This assignment of factors follows the convention that the fermion mass is measured in the units of the spatial lattice spacing,
$a_\sigma m$.
The dimensionless form of the action is then
\begin{equation}
\label{eq_Sf}
S_f=\sum_{x,y}\bar{\psi}_x M_{xy}\psi_y
=\sum_{x,y}\bar{\psi}_x\left[D_{xy,\sigma}+\xi_0^f D_{xy,\tau}+
a_\sigma m\delta_{xy}\right]\psi_y.
\end{equation}
The replacement of $\xi$ with $\xi_0^f$ indicates that the quantity in the action is a bare parameter (subscript ``0''), not necessarily equal to the target renormalized anisotropy $\xi$. Also, given that the discretization effects in the gauge and fermion sectors are of a different origin, this parameter is different (superscript ``$f$'') from the bare gauge anisotropy $\xi_0$.

To relate $\xi_0^f$ to the renormalized anisotropy, one needs to use observables sensitive to the fermion anisotropy. There are two common schemes:
\begin{itemize}
\item measure a state (\textit{e.g.}, a meson) at non-zero momentum $\vec{p}$ and fit the dispersion relation:
\begin{equation}\label{eq_disprel}
a_\tau^2E^2(p^2)=a_\tau^2E^2(0)+\frac{a_\sigma^2p^2}{(\xi^f)^2},
\end{equation}
\item measure a zero-momentum correlation function for, \textit{e.g.}, a meson with mass $M$, in the temporal and spatial directions and define the renormalized fermion anisotropy through the ratio of the extracted energy levels $E_\sigma$ and $E_\tau$
\begin{equation}\label{eq_effmassratio}
\xi^f=\frac{E_\sigma}{E_\tau}=\frac{a_\sigma M}{a_\tau M}=\frac{a_\sigma}{a_\tau}.
\end{equation}
\end{itemize}
The bare quantity $\xi_0^f$ is adjusted until $\xi^f=\xi$, the target renormalized anisotropy. We experimented with both approaches and prefer the first one. Our numerical setup is discussed in detail in Sec.~\ref{sec_fermion_ani}. The second approach is illustrated in Appendix~\ref{sec_app_mt_mz}.

So far the discussion has been generic and applies to any unsmeared Dirac operator. 
The spin Clifford algebra is realized for any staggered fermion formulation via phase factors $\eta_{x,\mu}$ accompanying either forward or backward displacements by an odd number of steps $\delta_{x\pm(2n+1)\hat\mu,y},\,n\in \mathbb{Z}$, together with some realization of the necessary parallel transporters $X_{x,\mu}[U]$, 
\begin{equation}\label{eq_spin_Ga}
S^\gamma_{\mu}(x,y) \equiv \eta_{x,\mu} X_{x,\mu}[U] \delta_{x\pm\hat\mu,y}
,\quad
\eta_{x,\mu} \equiv (-1)^{\sum\limits_{\nu<\mu} x_\nu}.
\end{equation} 
For the highly improved staggered quark action (HISQ)~\cite{Follana:2006rc} the Dirac operator has the following form (we follow the notation introduced in Ref.~\cite{MILC:2010pul}):
\begin{eqnarray}\label{eq_hisq}
2D^{HISQ}_{xy}[U]&=&\sum_{\mu=1}^4
\eta_{x,\mu}
\left\{
X_{x,\mu}^F[U]\delta_{x+\hat\mu,y}
-
X_{x-\hat\mu,\mu}^{F\dagger}[U]\delta_{x-\hat\mu,y}
\right.
\nonumber\\
&+&
\left.
(1+\epsilon(am))
\left(
X_{x,\mu}^L[U]\delta_{x+3\hat\mu,y}
-
X_{x-3\hat\mu,\mu}^{L\dagger}[U]\delta_{x-3\hat\mu,y}
\right)
\right\}
\end{eqnarray}
where the smeared links are constructed as $U_{x,\mu}\xrightarrow[]{Fat7}V_{x,\mu}$, 
$V_{x,\mu}\xrightarrow[]{proj~U(3)}W_{x,\mu}$, 
$W_{x,\mu}\xrightarrow[]{Fat7}X^F_{x,\mu}$, and
$W_{x,\mu}\xrightarrow[]{Naik}X^L_{x,\mu}$.
As the HISQ action involves smearing, there is potentially some freedom in where the anisotropy can be introduced: before or after the smearing. However, as the Fat7 part of the smearing was designed to suppress the coupling to gluons with momenta at the corners of the Brillouin zone~\cite{Blum:1996uf,Orginos:1999cr},
the relative coefficients of the Fat3, Fat5 and Fat7 paths are fixed.
Thus,
to preserve that property (which is independent of anisotropy),
we must introduce the fermion anisotropy at the outermost level:
\begin{eqnarray}
\label{eq_ahisq}
2D^{aHISQ}_{xy}[U]&=&\sum_{\mu=1}^3
\eta_{x,\mu}
\left\{
X_{x,\mu}^F[U]\delta_{x+\hat\mu,y}
-
X_{x-\hat\mu,\mu}^{F\dagger}[U]\delta_{x-\hat\mu,y}
\right.
\nonumber\\
&+&
\left.
(1+\epsilon(a_\sigma m))
\left(
X_{x,\mu}^L[U]\delta_{x+3\hat\mu,y}
-
X_{x-3\hat\mu,\mu}^{L\dagger}[U]\delta_{x-3\hat\mu,y}
\right)
\right\}
\nonumber\\
&+&\xi_0^f\eta_{x,4}
\left\{
X_{x,4}^F[U]\delta_{x+\hat4,y}
-
X_{x-\hat4,4}^{F\dagger}[U]\delta_{x-\hat4,y}
\right.
\nonumber\\
&+&
\left.
(1+\epsilon(a_\sigma m))
\left(
X_{x,4}^L[U]\delta_{x+3\hat4,y}
-
X_{x-3\hat4,\mu}^{L\dagger}[U]\delta_{x-3\hat4,y}
\right)
\right\}.
\end{eqnarray}
We call the action in Eq.~(\ref{eq_ahisq}) the anisotropic highly improved staggered quark (aHISQ) action.

This way of introducing the fermion anisotropy is also less demanding from the computational perspective, as the smearing and the major part of the fermion force, discussed in detail in Sec.~\ref{sec_dyn_ani} proceed in the same way for isotropic and anisotropic simulations.

To better understand the taste symmetry breaking pattern and to crosscheck our code, we have also performed calculations with naive (\textit{i.e.}, unimproved) staggered fermions with the following anisotropic Dirac operator
\begin{equation}
\label{eq_naive}
2D^{naive}_{xy}[U]=\sum_{\mu=1}^3
\eta_{x,\mu}
\left\{
U_{x,\mu}\delta_{x+\hat\mu,y}
-
U_{x-\hat\mu,\mu}^{\dagger}\delta_{x-\hat\mu,y}
\right\}
+\xi_0^f\eta_{x,4}
\left\{
U_{x,4\vphantom{\hat 4}}\delta_{x+\hat4,y}
-
U_{x-\hat4,4}^{\dagger}\delta_{x-\hat4,y}
\right\}.
\end{equation}
It is structurally the same as aHISQ, apart from the absence of the three-link hopping (Naik) term and the fact that the original gauge links $U_{x,\mu}$ are used.

\subsection{Properties of the anisotropic staggered spectrum}
\label{sec_ani_stag_spec}

Another Clifford algebra besides the one generated by Eq.~\eqref{eq_spin_Ga} for the spin exists among the internal taste degrees of freedom. 
It is realized similarly via a different set of phase factors $\zeta_{x,\mu}$~\cite{vandenDoel:1983mf,Golterman:1984cy}:
\begin{equation}\label{eq_taste_Xi}
	S^\xi_{\mu}(x,y) \equiv \zeta_{x,\mu} X_{x,\mu}[U] \delta_{x\pm\hat\mu,y}
	,\quad
	\zeta_{x,\mu} \equiv (-1)^{\sum\limits_{\nu>\mu} x_\nu}
	,
\end{equation} 
that implement the shift invariance of staggered fermion actions~\cite{Golterman:1984cy}. 
In the free theory, the products of phases and displacements of either kind constitute two separate exact Clifford algebras, \textit{e.g.},  
\begin{eqnarray}\label{eq_spin_CA}
	S^\gamma_{\mu}(x,y) S^\gamma_{\nu}(y,z) 
	+ S^\gamma_{\nu}(x,y) S^\gamma_{\mu}(y,z) 
	&=& 2\delta_{\mu\nu} \delta_{x\pm\hat\mu\pm\hat\nu,z},\\
	\label{eq_taste_CA}
	S^\xi_{\mu}(x,y) S^\xi_{\nu}(y,z) 
	+ S^\xi_{\nu}(x,y) S^\xi_{\mu}(y,z) 
	&=& 2\delta_{\mu\nu} \delta_{x\pm\hat\mu\pm\hat\nu,z}
\end{eqnarray}
up to an overall two-step displacement, and elements from separate spin or taste Clifford algebras commute. 
Thus, Eq.~\eqref{eq_taste_CA} enforces exact sixteenfold degeneracies for the free staggered spectrum. 

Yet in the interacting theory the displacements $S^\gamma,S^\xi$ in Eqs.~\eqref{eq_spin_Ga} and \eqref{eq_taste_Xi} are accompanied by parallel transporters $X_{x,\mu}[U]$ that do not commute. 
Hence, Wilson loop factors arise between the two paths in Eqs.~\eqref{eq_spin_CA} and~\eqref{eq_taste_CA} (if one factors out a common part of the paths), which cause mixing and lift degeneracies in the interacting staggered spectrum. 
These outcomes are irrespective of using forward or backward displacements, or of any particular choices for the parallel transporters. 

Site-local operators are constructed by pairing corresponding elements of both Clifford algebras with one forward and one backward displacement, e.g. $S^\gamma_{\mu}(x,x\pm\hat\mu)S^\xi_{\mu}(x\mp\hat\mu,x)$, most notably in the spin-pseudoscalar taste-pseudoscalar case, 
\begin{equation}\label{eq_eps}
	\epsilon_{x} \equiv S^\gamma_{5}(x,y) S^\xi_{5}(y,x) = (-1)^{\sum\limits_{\mu} x_\mu}
	,
\end{equation}
which is the generator of the last remaining taste-isovector $U(1)_A$ chiral symmetry of any staggered fermion action. 
Because this symmetry is broken spontaneously in the interacting theory, the lowest pseudoscalar taste state excited by interpolating operators using $\epsilon_{x}$ is a Goldstone boson. 

Any other taste of a pseudoscalar state needs further parallel transport on the taste end, by one step in the $\mu$ direction for $S^\xi_{5}(y,x)S^\xi_{\mu}(x,x\pm\hat\mu)$, by one step each in the $\mu$ and $\nu$ directions for $S^\xi_{5}(y,x) S^\xi_{\mu}(x,x\pm\hat\mu) S^\xi_{\nu}(x\pm\hat\mu,x\pm\hat\mu\pm\hat\nu)$, etc. 
Each of those additional steps involves gauge links that fluctuate at the lattice scale, less so if smeared. 
Their fluctuations, which are independent to a leading approximation, add positive contributions scaling as $(a\Lambda)^2$, where $\Lambda$ represents a typical hadronic scale, to the respective taste hadron's squared energy. 
As such, the sixteenfold degeneracy breaks up in the interacting theory, and the spectrum is affected by taste-symmetry breaking.  
In improved staggered quark formulations using smoother links such as HISQ, \textit{i.e.}, Eq.~\eqref{eq_hisq}, the link fluctuations are reduced, and thus the taste-symmetry breaking is diminished~\cite{Blum:1996uf, Orginos:1999cr, Follana:2006rc}. 

With respect to the taste quantum number, the meson multiplets are classified as pseudoscalar (denoted $\xi_5$ or P), axial vector ($\xi_\mu\xi_5$, A), tensor ($\xi_\mu\xi_\nu$, T), vector ($\xi_\mu$, V) and scalar, or singlet (1, S)~\cite{Golterman:1985dz}, where $\xi_\mu$ are the Euclidean Dirac $\gamma$-matrices, denoted $\xi_\mu$ to distinguish the taste from the spin basis.
A taste-multiplet hierarchy of $P<A<T<V<S$ with $1,4,6,4,1$-fold remnant degeneracies emerges for the tastes of pseudoscalar states due to the minimal count of fluctuating links involved in each of the interpolating operators. 
The Goldstone boson taste $P$ is the lowest and the taste singlet pseudoscalar boson taste $S$ is the highest state. 
Since the pion mass is much smaller than the hadronic scale $\Lambda$, taste-symmetry breaking is a large effect for the taste pions. 
Because all other light hadrons have masses at the hadronic scale $\Lambda$, taste-symmetry breaking is only a modest effect for those. 

On anisotropic lattices, the temporal links are smoother than the spatial links, and hence, the former fluctuate less than the latter. 
Thus, we are led to expect that a new taste-multiplet hierarchy
$P\lesssim A_{\tau} < A_{\sigma} \lesssim T_{\sigma\sigma} < T_{\sigma\tau} \lesssim V_{\sigma} < V_{\tau} \lesssim S$ with $1,1,3,3,3,3,1,1$-fold remnant degeneracies emerges at nontrivial anisotropy. 
In the limit of infinite anisotropy, \textit{i.e.}, vanishing temporal lattice spacing, the fluctuations of the temporal links become negligible, the taste matrix $S^\xi_{4}(x,x\pm\hat4)$ does not change the energy anymore, and the approximate $\lesssim$ degeneracies become exact with a $2,6,6,2$-fold degeneracy pattern.

\section{Dynamical fermions with anisotropy}
\label{sec_dyn_ani}

As in the isotropic dynamical HISQ simulations, we intend to use the Rational Hybrid Monte Carlo (RHMC)
algorithm~\cite{Clark:2003na,Clark:2006fx,Clark:2006wp}. The main components of RHMC are the gauge and fermion forces, and the fermion matrix inverter. For aHISQ we an need anisotropic version of all the components. The gauge force is straightforward, and it was developed and tested in parallel with the anisotropic local updating algorithms for pure gauge ensembles that are discussed in Sec.~\ref{sec_pg_ens}.

The aHISQ inverter amounts to introducing proper anisotropy factors as shown in Eq.~(\ref{eq_Sf}). Those can be taken care of by multiplying the HISQ smeared links immediately before the inversion. The conjugate gradient proceeds in exactly same way as for the isotropic case. The most complicated component is the fermion force: the anisotropy factors are on the outermost smearing level, however, the chain rule proceeds recursively outside-in. The anisotropy factors can be incorporated at the stage of constructing the outer products of pseudo-fermion fields. Structurally, the aHISQ fermion force is similar to HISQ, therefore we follow Ref.~\cite{Wong:2007uz} that spelled out the recursive chain rule expressions for multiply smeared actions, such as HISQ.

To simplify the discussion, it is enough to consider a single-smeared action, for instance, Fat7~\cite{Blum:1996uf,Orginos:1999cr}: $U_{x,\mu}\xrightarrow[]{Fat7}V_{x,\mu}$. We closely follow the discussion of Ref.~\cite{Wong:2007uz} in this part. The smeared links $V_{x,\mu}$ are constructed as weighted averages of the products of the original gauge links $U_{x,\mu}$ along the paths of length up to 7. The Dirac operator for the anisotropic Fat7 action is
\begin{equation}
\label{eq_smeared_diracop}
D_{xy}=\sum_{\mu=1}^3\eta_{x,\mu}
\left\{V_{x,\mu}\delta_{x,y-\hat\mu}-V^\dagger_{x-\hat\mu,\mu}\delta_{x,y+\hat\mu}\right\}
+\xi_0^f\ \eta_{x,4}
\left\{
V_{x,4}\delta_{x,y-\hat4}-V^\dagger_{x-\hat4,4}\delta_{x,y+\hat4}
\right\}
\end{equation}
and the fermion matrix is $M_{xy}=D_{xy}+2m\delta_{xy}$ (for convenience in the code $M$ and $D$ are redefined to be twice what they are in Eq.~(\ref{eq_Sf})).

For $n_f$ fermion flavors the rooted staggered fermion action is
\begin{equation}
\label{eq_Sf_phi}
S_f=\langle\Phi|[M^\dagger M]^{-n_f/4}|\Phi\rangle.
\end{equation}
In the RHMC algorithm the fractional power is approximated as
\begin{equation}
[M^\dagger M]^{-n_f/4}\approx \alpha_0+\sum_l \frac{\alpha_l}{M^\dagger M+\beta_l}
\end{equation}
and the fermion force is
\begin{equation}
\label{eq_fxmu_def}
f_{x,\mu}=\frac{\delta S_f}{\delta U_{x,\mu}}=-\sum_l\alpha_l\left\{
\left\langle X^l\left|\frac{\partial D^\dagger[V[U]]}{\partial U_{x,\mu}}\right|Y^l\right\rangle
+
\left\langle Y^l\left|\frac{\partial D[V[U]]}{\partial U_{x,\mu}}\right|X^l\right\rangle
\right\},
\end{equation}
where $|X^l\rangle=[M^\dagger M+\beta_l]^{-1}|\Phi\rangle$ and $|Y^l\rangle=D|X^l\rangle$. The $X^l$ and $Y^l$ fields are defined on the even and odd sites, respectively, and they implicitly incorporate the anisotropy, as they are solutions to a linear system with the anisotropic Dirac operator. Using the chain rule,
\begin{eqnarray}
\frac{\partial D_{mn}}{\partial U_{x,\mu}}&=&
\sum_{y,\nu}\bigg[\frac{\partial D_{mn}}{\partial V_{y,\nu}}\frac{\partial V_{y,\nu}}{\partial U_{x,\mu}}+\frac{\partial D_{mn}}{\partial V^\dagger_{y,\nu}}\frac{\partial V^\dagger_{y,\nu}}{\partial U_{x,\mu}}\bigg],\nonumber\\
\frac{\partial D^\dagger_{mn}}{\partial U_{x,\mu}}&=&
\sum_{y,\nu}\bigg[\frac{\partial D^\dagger_{mn}}{\partial V_{y,\nu}}\frac{\partial V_{y,\nu}}{\partial U_{x,\mu}}+\frac{\partial D^\dagger_{mn}}{\partial V^\dagger_{y,\nu}}\frac{\partial V^\dagger_{y,\nu}}{\partial U_{x,\mu}}\bigg],
\label{eq_dirac0}
\end{eqnarray}
where the derivatives $\partial V/\partial U$ depend only on the smearing and are not affected by the anisotropy. The other derivatives involve the anisotropy directly:
\begin{eqnarray}
\frac{\partial D_{mn}}{\partial V_{y,\nu}}&=&\omega_\nu\eta_{m,\nu}\delta_{m,y}\delta_{m,n-\hat\nu},\nonumber\\
\frac{\partial D_{mn}}{\partial V^\dagger_{y,\nu}}&=&-\omega_\nu\eta_{m,\nu}\delta_{m-\hat\nu,y}\delta_{m,n+\hat\nu},\nonumber\\
\frac{\partial D^\dagger_{mn}}{\partial V_{y,\nu}}&=&-\omega_\nu\eta_{n,\nu}\delta_{n-\hat\nu,y}\delta_{n,m+\hat\nu},\nonumber\\
\frac{\partial D^\dagger_{mn}}{\partial V^\dagger_{y,\nu}}&=&\omega_\nu\eta_{n,\nu}\delta_{n,y}\delta_{n,m-\hat\nu},
\label{eq_dirac_ani}
\end{eqnarray}
with
\begin{equation}
\omega_\nu=\left\{
\begin{array}{ll}
1, & \nu\neq4,\\
\xi_0^f, & \nu=4.
\end{array}
\right.
\end{equation}
Substituting Eqs.~(\ref{eq_dirac0}) and (\ref{eq_dirac_ani}) into Eq.~(\ref{eq_fxmu_def}) gives
\begin{eqnarray}
[f_{x,\mu}]_{AB}&=&\sum_{y}(-1)^y
\bigg[\sum_{\nu\neq4}\eta_{y,\nu}\Big(\frac{\partial [V_{y,\nu}]_{CD}}{\partial [U_{x,\mu}]_{AB}}[f^{(0)}_{y,\nu}]_{CD}+\frac{\partial [V^{\dagger}_{y,\nu}]_{CD}}{\partial [U_{x,\mu}]_{AB}}[f^{(0)\dagger}_{y,\nu}]_{CD}\Big)\nonumber\\
&+&\xi_0^f\ \eta_{y,4}\Big(\frac{\partial [V_{y,4}]_{CD}}{\partial [U_{x,\mu}]_{AB}}[f^{(0)}_{y,4}]_{CD}+\frac{\partial [V^{\dagger}_{y,4}]_{CD}}{\partial [U_{x,\mu}]_{AB}}[f^{(0)\dagger}_{y,4}]_{CD}\Big)\bigg],\label{eq_fmu}
\end{eqnarray}
which is an anisotropic generalization of Eq.~(2.6) of Ref.~\cite{Wong:2007uz}, and where
\begin{equation}
\label{eq_f0_def}
[f^{(0)}_{y,\nu}]_{CD}=\sum_{l}\alpha_l\big([Y^l_{y+\nu}]_D[X^{l*}_y]_C+[X^l_{y+\nu}]_D[Y^{l*}_y]_C\big).
\end{equation}
The color indices are explicitly spelled out with capital roman letters.

The bare fermion anisotropy $\xi_0^f$ multiplies the temporal component of $f^{(0)}_{x,\mu}$ and this suggests an economical practical implementation: multiply the outer product $|X\rangle\langle Y|$ by $\xi_0^f$ when the corresponding sites are separated by a temporal link at the stage of constructing $f^{(0)}_{x,\mu}$. After that, the chain rule proceeds identically for the isotropic and anisotropic case.

\section{Numerical results}
\label{sec_num_res}
To validate our code for anisotropic simulations, understand anisotropy tuning with the gradient flow and study the spectrum of anisotropic staggered quarks, we start with anisotropic pure gauge simulations. In the fully dynamical case that we are interested in, \textit{i.e.}, two degenerate light and one strange quark, one needs to tune four bare parameters simultaneously: the gauge coupling $\beta$, gauge anisotropy $\xi_0$, strange quark mass $m_s$ (as is often done, we can fix the strange-to-light mass ratio, \textit{e.g.}, to $m_s/m_l=5$), and the fermion anisotropy $\xi_0^f$. For quenched simulations the tuning splits into two independent steps: $(\beta,\xi_0)$, then $(m_s,\xi_0^f)$.

\subsection{Pure gauge ensembles}
\label{sec_pg_ens}

We anticipate using ensembles with high anisotropy with Non-Relativistic QCD (NRQCD) formalism for heavy quarks. Therefore we are interested in relatively spatially coarse lattices. To keep the computational cost of the explorations modest, we start with the spatial lattice spacing $a_\sigma\sim 0.16-0.2$~fm and concentrate our simulations in that region of the parameter space to understand how the discretization effects influence anisotropy tuning and what anisotropies can practically be reached. (As is clear from Eq.~(\ref{eq_gauge_action}), higher anisotropy is equivalent to stronger coupling in the magnetic part of the action.)

To generate anisotropic pure gauge ensembles we use the Cabibbo-Marinari SU(3) heatbath algorithm~\cite{Cabibbo:1982zn} that updates the three SU(2) subgroups. The SU(2) matrices are sampled with the Fabricius-Haan-Kennedy-Pendleton (FHKP) heatbath~\cite{Fabricius:1984wp,Kennedy:1985nu}. Per one step of the heatbath algorithm, four steps of overrelaxation \cite{Adler:1987ce} were performed. We also generated several test ensembles with the Hybrid Monte Carlo algorithm~\cite{Duane:1987de} to validate the anisotropic gauge force. 
Among a number of tests, the anisotropic pure gauge code was validated by reproducing some runs from Ref.~\cite{CP-PACS:2001lwl}, as discussed in Appendix~\ref{sec_app_tests}.

\subsection{Gauge anisotropy and lattice spacing tuning}

To determine the spatial and temporal lattice spacing, $(a_\sigma,a_\tau)$, or, equivalently, the spatial lattice spacing and renormalized anisotropy, $(a_\sigma,\xi)$, we use the gradient flow as described in Sec.~\ref{sec_ani_flow}. The discretization effects on the scale setting come from three origins: the gauge action used to generate the configurations, generator of the flow and the discretization of the observable, as is well known for the isotropic case~\cite{Fodor:2014cpa}.

The gauge action is fixed to the tree-level Symanzik-improved one, Eq.~(\ref{eq_gauge_action}). For the flow we use Wilson (W), tree-level Symanzik (S) action, and, in some cases, we also compute the Zeuthen flow (Z), which is a modification of the flow equation itself~\cite{Ramos:2015baa}. For the observable we use Wilson (W), Symanzik (S), clover (C)~\cite{Sheikholeslami:1985ij} and improved clover (I)~\cite{Bilson-Thompson:2002xlt} discretization. As the gauge action is fixed, we only denote flow-observable combinations with two-letter abbreviations, spanning [W,S,Z][W,S,C,I].
The early flow time artifacts are different for these combinations, and their severity limits how large the spatial lattice spacing may be so that the tuning procedure of Sec.~\ref{sec_ani_flow} is still practical to use.

\begin{figure}
\includegraphics[width=0.49\textwidth]{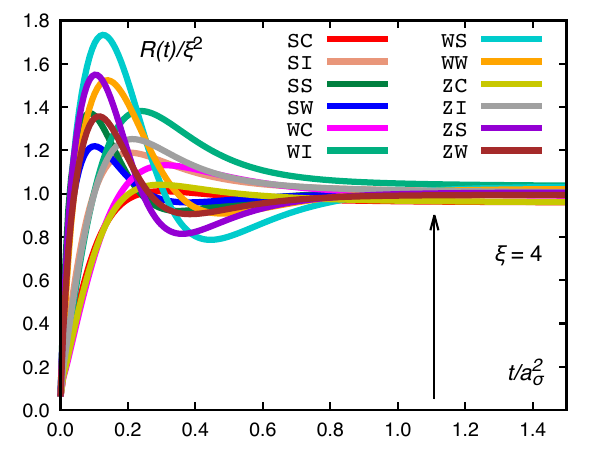}\hfill
\includegraphics[width=0.49\textwidth]{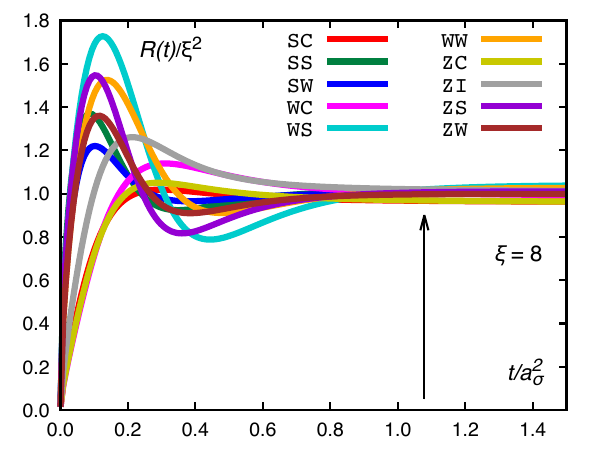}
\caption{The ratio defined in Eq.~(\ref{eq_Rt}) scaled with the renormalized anisotropy, $R(t)/\xi^2$, plotted as a function of flow time $t/a_\sigma^2$ for the renormalized anisotropy $\xi=4$ (left) and $\xi=8$ (right). The vertical lines correspond to $t=w_{0,\sigma}^2/a_\sigma^2$ with $a_\sigma\approx0.1665$~fm. The lines of different color correspond to different flow-observable schemes. The errorbars are not visible on the scale of the figure. \label{fig_flows}}
\end{figure}

To assess the discretization effects on the anisotropy, which is determined through the ratio in Eq.~(\ref{eq_tune2}), one can compute the evolution of the ratio of the energy densities defined in Eq.~(\ref{eq_tune1}) themselves:
\begin{equation}
\label{eq_Rt}
R(t)=t\frac{d}{dt} t^2 \langle S_{\sigma\sigma}(t) \rangle\left/\ t\frac{d}{dt} t^2 \langle S_{\sigma\tau}(t) \rangle\right..
\end{equation}
When $w_{0,\sigma}/w_{0,\tau}$ is close to 1, $R(t)$ approaches $\xi^2$. In Fig.~\ref{fig_flows} the ratio $R(t)/\xi^2$ is shown for the two tuned ensembles that we use in this study, with same spatial lattice spacing $a_\sigma\approx0.1665$~fm and the renormalized gauge anisotropy $\xi=4$ (left panel) and $\xi=8$ (right panel). The first important point the plots illustrate is that, as expected, at large anisotropies the discretization effects are determined by the spatial lattice spacing $a_\sigma$: the ratios $R(t)/\xi^2$ for the two ensembles behave almost identically. We have not measured the SI and WI combinations on the $\xi=8$ ensemble (right panel), but we expect them also to be similar to the ones at $\xi=4$ (left panel). The arrows indicate the flow time that corresponds to $a_\sigma\approx0.1665$~fm. This flow time is in the flat region of $R(t)$, however, with NRQCD and thermodynamics applications in mind, we are interested in even coarser lattices.
Therefore for tuning the gauge anisotropy we prefer such flow-observable combinations that quickly reach plateau and are monotonic for the most part. From Fig.~\ref{fig_flows} one can observe that, in general, the clover (C) and improved clover (I) observables reach a maximum and then approach the plateau from above, while Wilson (W) and Symanzik (S) observables reach maximum, then minimum and then approach the plateau from below. We therefore avoid using W and S observables for tuning the gauge anisotropy. Another observation from the left panel of Fig.~\ref{fig_flows} is the ordering of the combinations of W, S, Z flows and C, I observables from the least to most variation (\textit{i.e.}, the peak height): SC, ZC, WC, SI, ZI, WI. Given that ordering, we will choose the clover observable (C) for tuning.

\begin{figure}
\includegraphics[width=0.49\textwidth]{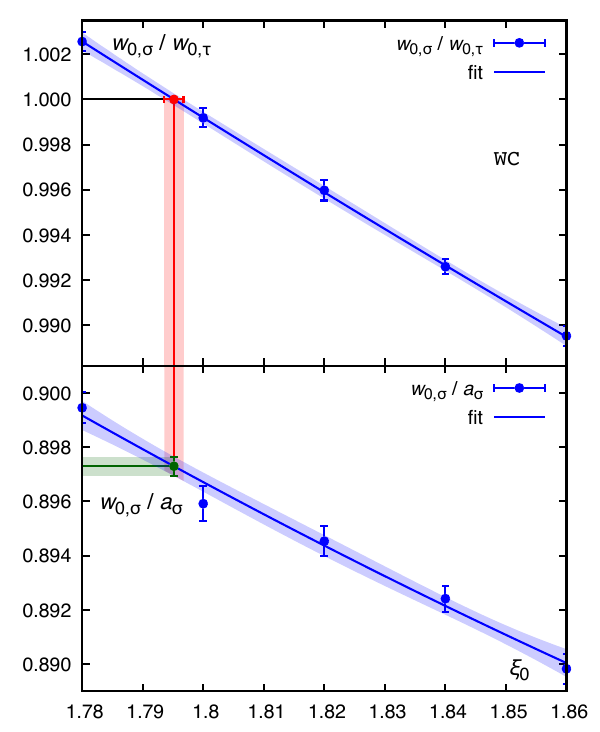}\hfill
\includegraphics[width=0.49\textwidth]{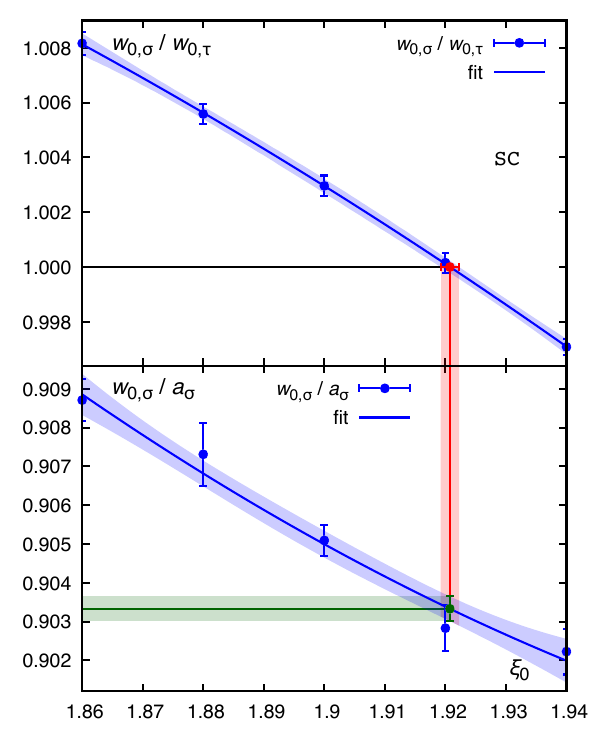}
\caption{Determining the bare gauge anisotropy $\xi_0$ and the lattice scale in units of $w_{0,\sigma}/a_\sigma$ for a pure gauge ensemble at $\beta=6.9$ and the renormalized anisotropy $\xi=2$ in the WC (left) and SC (right) flow-observable scheme.\label{fig_xig_tune}}
\end{figure}

Our discussion in this part follows closely the tuning process described in Ref.~\cite{Borsanyi:2018srz}.
The continuum limit is usually taken at fixed renormalized anisotropy. In the parameter space $(\beta,\xi_0)$ the fixed renormalized anisotropy $\xi(\beta,\xi_0)={\rm const}$ translates into an implicit dependence $\xi_0(\beta)$ which we call the line of constant renormalized anisotropy (LCRA). We now exemplify the tuning process for the $\xi=2$ LCRA. We generate ensembles on the grid that spans $\beta=\{6.9,...,7.3\}$ with step of 0.1 and $\xi_0=\{1.78,...,1.94\}$ with step of $0.02$, \textit{i.e.}, $5\times9=45$ pure gauge tuning ensembles with the same lattice volume $16^3\times32$ (a large enough physical volume to accommodate the lowest glueball and to allow for reaching the flow times on the scale of $w_0^2$ without distortions from the boundary). The large number of $\xi_0$ points is needed for studying different flow-observable combinations, as will become clear in a moment. Consider now fixed $\beta=6.9$ set of ensembles as an example. At each $\xi_0$ we compute W, S and Z flows with the flow anisotropy set to the target one, $\xi_0^{gf}=\xi=2$. For the WC combination the ratio $w_{0,\sigma}/w_{0,\tau}$ as function of the bare gauge anisotropy $\xi_0$ is shown in the top panel of the left panel of Fig.~\ref{fig_xig_tune}. We fit the dependence over the five nearby $\xi_0$ points with a linear or quadratic polynomial and propagate errors with bootstrap. The errors are uncorrelated -- each point represents an independent ensemble. Determining $\xi_0^*$ such that $w_{0,\sigma}(\xi_0^*)/w_{0,\tau}(\xi_0^*)=1$ gives the bare gauge anisotropy for which at $\beta=6.9$ the renormalized gauge anisotropy defined in the WC scheme is $\xi=2$. In this particular case, $\xi_0^*=1.7952(16)$. We similarly fit $w_{0,\sigma}(\xi_0)/a_\sigma$, shown in the bottom panel of the left panel of Fig.~\ref{fig_xig_tune}, and evaluate $w_{0,\sigma}(\xi_0^*)/a_\sigma=0.89729(36)$. Thus, from this set of the tuning runs we learn that an ensemble generated at $(\beta=6.9,\xi_0=1.7952)$ will have the target renormalized anisotropy $\xi=2$ and the gradient flow scale $w_{0,\sigma}/a_\sigma$ of about 0.897.

The right panel of Fig.~\ref{fig_xig_tune} illustrates the same process for $\beta=6.9$ and the SC scheme. In that case, $\xi_0^*=1.9207(15)$ and $w_{0,\sigma}(\xi_0^*)/a_\sigma=0.90333(32)$. As expected, the discretization effects lead to quite different values of $\xi_0$ for the two gradient flow schemes. 
For the production ensembles we will use only one scheme, for which we simply repeat the process described above for the sets of ensembles at different $\beta$ values. This produces $\xi=2$ LCRA encoded in the dependencies $\xi_0^*(\beta)$ and $w_{0,\sigma}(\xi_0^*,\beta)/a_\sigma$. To simplify the notation, we drop the asterisk $*$ and the $\xi_0^*$ dependence in the scale, implying that $\xi_0(\beta)$ and $w_{0,\sigma}(\beta)/a_\sigma$ represent LCRA that maps $(\beta,\xi_0(\beta))$ to the target $(a_\sigma=w_0/(w_{0,\sigma}(\beta)/a_\sigma),\xi)$.

To understand the discretization effects further, we measured multiple flow-observable combinations for $\xi=2$, and also repeated the same tuning process for $\xi=4$ LCRAs in multiple gradient flow schemes. In the latter case, the tuning ensembles spanned the same $\beta$ values, and the bare gauge anisotropies are $\xi_0=\{3.0,...,3.9\}$ with step of $0.1$. The lattice volume is $16^3\times64$.

\begin{figure}
\includegraphics[width=0.49\textwidth]{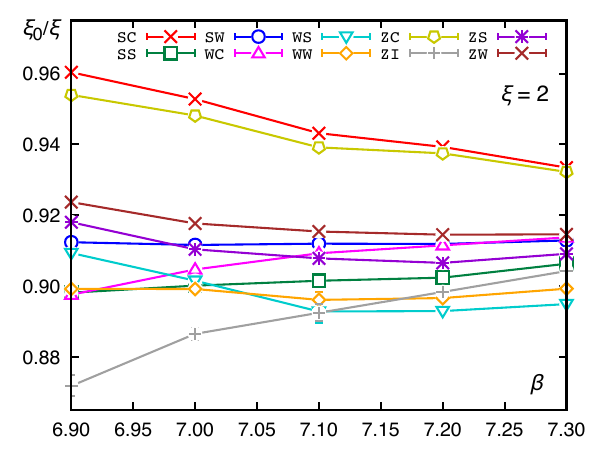}\hfill
\includegraphics[width=0.49\textwidth]{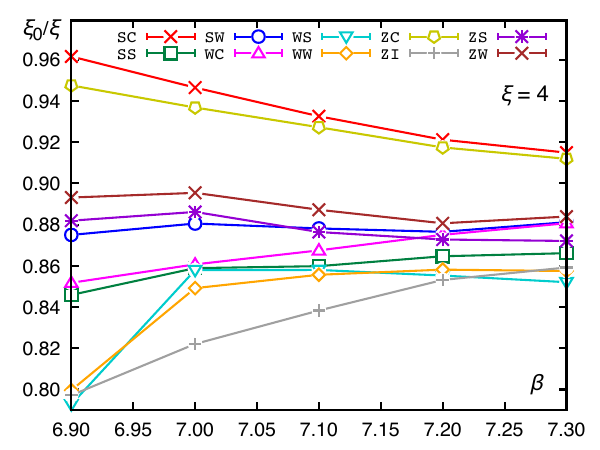}
\caption{The lines of constant renormalized anisotropy, \textit{i.e.}, functional dependencies $\xi_0(\beta)$ that correspond to the renormalized gauge anisotropy $\xi=2$ (left) and $\xi=4$ (right) in different flow-observable schemes.\label{fig_xig_schemes}}
\end{figure}
The resulting $\xi=2$ and $\xi=4$ LCRAs in different gradient flow schemes are shown in the left and right panels of Fig.~\ref{fig_xig_schemes}, respectively. The lines are drawn to simply guide the eye and are not fits. The errorbars are not visible on the scale of the figure.
For easier comparison, the $y$-axes are $\xi_0(\beta)/\xi$ rather than $\xi_0(\beta)$. In the limit $\beta\to\infty$ the ratio of the bare to renormalized anisotropy must approach $\xi_0/\xi\to1$.
The largest ratio is provided by the SC scheme and it may be preferred as the least amount of renormalization is needed.
However, the SC LCRA $\xi_0(\beta)$, and similarly the nearby ZC LCRA, is decreasing in the relevant $\beta$ range, which means that at some value of $\beta$ it must acquire a minimum and then approach 1 from below. This non-monotonicity of the SC LCRA most likely translates into the non-monotonicity in the approach to the continuum limit in the observables, if the simulations are tuned in the SC scheme. We, in general, would like to avoid such behavior. The more promising schemes are WC and ZI, and, perhaps, SS and SW. The WS, ZS, ZW exhibit a decrease similar to SC and ZC. WS and WW appear relatively flat for $\xi=2$ but exhibit a rapid drop at the lowest coupling $\beta=6.9$ for $\xi=4$ (Fig.~\ref{fig_xig_schemes}, right). We attribute that behavior to the early-flow-time artifacts, similar to the ones shown in Fig.~\ref{fig_flows}, since for $\xi=2$ LCRA $\beta=6.9$ corresponds to $a_\sigma\approx0.20$~fm, while for $\xi=4$ LCRA to $a_\sigma\approx0.24$~fm. The two schemes that exhibit monotonically rising behavior are WC and ZI. The ZI LCRA is significantly below the WC LCRA, meaning that $\xi_0$ requires larger renormalization. We therefore use WC as the tuning scheme for all production ensembles.
With the WC tuning scheme and the choice of the gauge action in Eq.~(\ref{eq_gauge_action}) our pure gauge setup is identical to the one of Ref.~\cite{Borsanyi:2018srz}, which used the renormalized anisotropy $\xi=2$.

\begin{figure}
\includegraphics[width=0.49\textwidth]{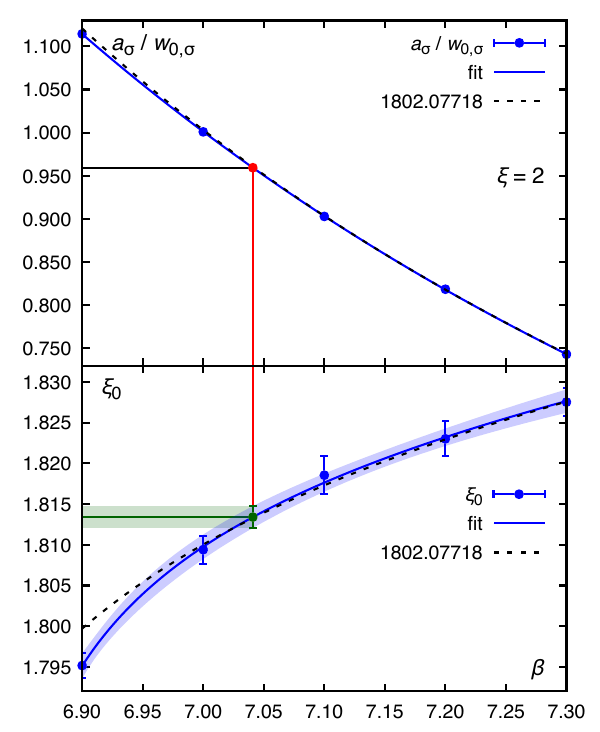}\hfill
\includegraphics[width=0.49\textwidth]{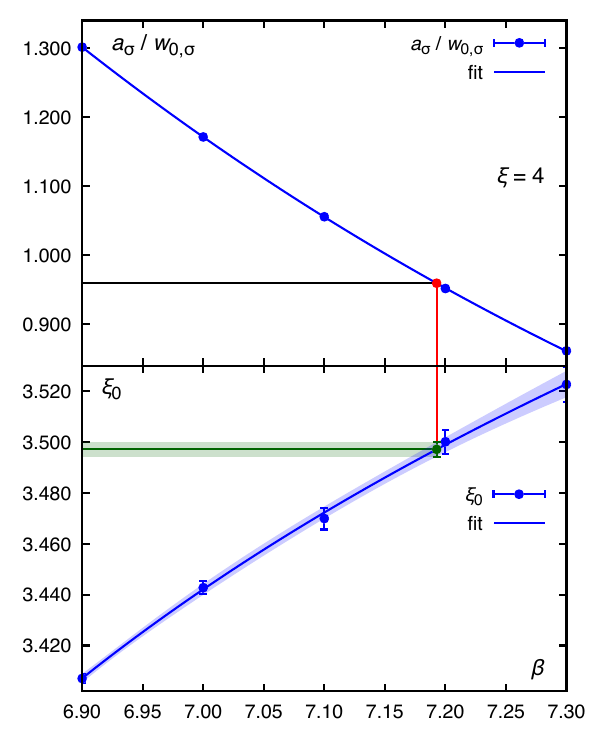}
\caption{Finding the bare gauge coupling $\beta$ (top panels) and the bare gauge anisotropy $\xi_0$ (bottom panels) that correspond to the spatial lattice spacing $a_\sigma=0.1665$~fm and the renormalized anisotropy $\xi=2$ (left panels) and 4 (right panels), in the WC scheme, as explained in the text. The curves in the bottom panels are the same as the magenta lines with the triangle symbols in Fig.~\ref{fig_xig_schemes}, shown as $\xi_0$ on the $y$-axis.}\label{fig_lcra}
\end{figure}

For parametrization purposes it is more convenient to work with the inverse gradient flow scale. The $\xi=2$ and $\xi=4$ LCRAs that we tuned in the range of bare gauge couplings $\beta=6.9,...,7.3$ are shown, respectively, in the left and right panels of Fig.~\ref{fig_lcra}. The top panels display $a_\sigma/w_{0,\sigma}(\beta)$ and the bottom $\xi_0(\beta)$. The dashed lines in the left panel show the corresponding parametrizations of the $\xi=2$ LCRA from Ref.~\cite{Borsanyi:2018srz}. The deviations for $\beta<7$ can be attributed to the fact that the lowest coupling considered there was $\beta=7$ ($\beta=4.2$ in their standard normalization $\beta=2N_c/g_0^2$) and thus the fits were not constrained below that point.

The inverse gradient flow scale is fit with the Allton-type ansatz~\cite{Allton:1996dn}:
\begin{equation}
\frac{a_\sigma}{w_0}(\beta)=c_0f(\beta)\frac{1+c_2(10/\beta)f(\beta)^2}
{1+d_2(10/\beta)f(\beta)^2},
\end{equation}
\begin{equation}
f(\beta)=\left(\frac{10b_0}{\beta}\right)^{-b_1/(2b_0^2)}
\exp\left(-\frac{\beta}{20b_0}\right),
\end{equation}
where the two-loop beta-function coefficients for the $N_c=3$ pure gauge case are $b_0=11N_c/(48\pi^2)$ and $b_1=34N_c^2/(768\pi^4)$. For the $\xi=2$ coefficients we find:
\begin{equation}
c_0=42.08(45),\,\,\,c_2=856(152),\,\,\,d_2=2.8(85.9),\nonumber
\end{equation}
and for $\xi=4$:
\begin{equation}
c_0=43.13(75),\,\,\,c_2=2539(255),\,\,\,d_2=792(108).\nonumber
\end{equation}
To fit $\xi_0(\beta)$ we use the same rational function ansatz as in Ref.~\cite{Borsanyi:2018srz} ($\xi$ is fixed to the target value and is not a fit parameter):
\begin{equation}
\label{eq_xi0beta_fit}
\xi_0(\beta)=\xi\left(1+\frac{10}{\beta}\,
\frac{u_2+u_4(10/\beta)}{1+v_2(10/\beta)}\right),
\end{equation}
where for $\xi=2$
\begin{equation}
u_2=-0.05958(66),\,\,\,u_4=0.03989(64),\,\,\,v_2=-0.6727(27),\nonumber
\end{equation}
and for $\xi=4$
\begin{equation}
u_2=-0.039(19),\,\,\,u_4=0.013(18),\,\,\,v_2=-0.562(47).\nonumber
\end{equation}
Negative $u_2$ reflects the fact that the approach to $\xi$ is from below. Converted to our convention, the corresponding coefficients of the fit 
form~(\ref{eq_xi0beta_fit}) for $\xi=2$ from Ref.~\cite{Borsanyi:2018srz} are: 
$u_2=-0.0578007$, $u_4=0.0375841$ and $v_2=0.65674$.

For the study of the fermion anisotropy tuning and properties of the staggered spectrum we  tuned ensembles with various gauge anisotropies and fixed spatial lattice spacing $a^*_\sigma\approx0.1665$~fm in the WC scheme. The use of the LCRA for tuning is also illustrated in Fig.~\ref{fig_lcra}. That lattice spacing corresponds to $a^*_\sigma/w_{0}=0.1665/0.17355\approx0.959378$. (We use $w_0=0.17355(92)$~fm from the FLAG Review 2024~\cite{FlavourLatticeAveragingGroupFLAG:2024oxs}.) Solving $a_\sigma/w_{0,\sigma}(\beta)=a^*_\sigma/w_{0}$ for $\beta$ from the top curve and then computing $\xi_0$ at that $\beta$ from the lower curve gives the needed parameters. For $\xi=2$ we have $\beta=7.0410(34)$ and $\xi_0=1.8134(13)$, and for $\xi=4$ $\beta=7.19278(39)$ and $\xi_0=3.4971(28)$.

In addition to $\xi=2$ and $\xi=4$ we also tuned a set of ensembles with the same spatial lattice spacing $a_\sigma\approx0.1665$~fm and the renormalized anisotropies $\xi=1$, $1.1$, $1.2$, $1.5$ and $8$. The parameters of the ensembles are summarized in Table~\ref{tab_pg_ens} in Appendix \ref{sec_app_ens}. The mistuning of anisotropy can be measured by deviation of the $w_{0,\sigma}/w_{0,\tau}$ ratio from one. The largest in our case is 0.1\% (for $\xi=8$). The largest deviation of the spatial lattice spacing from the target value $a_\sigma=0.1665$~fm is 
0.4\% (also for $\xi=8$). We find, expectedly, that the gradient flow anisotropy tuning procedure is statistically quite  accurate.

The gradient flow is integrated with the 6-stage 4-order low-storage Runge-Kutta scheme of Ref.~\cite{BERLAND20061459} which happens to be a low-storage commutator-free Lie group integrator, as explained in Refs.~\cite{Bazavov2021,BazavovChuna2021,Bazavov2026,Bazavov:2025exj}. The setup is identical to, \textit{e.g.}, the one used in Ref.~\cite{FermilabLattice:2025dui}, apart from the anisotropy. While in Ref.~\cite{FermilabLattice:2025dui} the integration was performed at step sizes as large as $1/20$, here one should take into account the potentially large anisotropy factors, Eq.~(\ref{eq_ani_factors}), that appear in the generator of the flow. For that reason, for instance, at the largest anisotropy $\xi=8$, we used the step size of $1/128$ to keep the integration stable.

\subsection{Meson correlation functions fits}

Our spectrum computational setup is very similar to the one used in Refs.~\cite{MILC:2010pul,MILC:2012znn}.
For tuning spectrum measurements with zero and finite momentum we used corner wall sources, and for pion splittings measurements we used the special wall sources that include phases and allow us to project different meson tastes at the sink, which are also standard in the MILC code. The meson correlation functions are fit to the expected staggered form that includes oscillating channel and the back-propagating states due to the finite temporal extent of the lattice:
\begin{equation}\label{eq_corr_fitform}
	C(\tau)=\sum_{i=1}^2A^n_i\big(e^{-E^n_i\tau}+e^{-E^n_i(N_\tau-\tau)}\big)+(-1)^\tau\sum_{j=1}^2A^o_j\big(e^{-E^o_j\tau}+e^{-E^o_j(N_\tau-\tau)}\big),
\end{equation}
where the superscripts ``$n$" and ``$o$" denote non-oscillating and oscillating states, respectively.

Where necessary, we include up to one excited state in each channel. The consistency of the fits is checked by comparing the results with and without the excited states and by varying the minimal distance included in the fit $\tau_{min}$. All fits whose results are listed in Tables~\ref{tab_tune_ahisq_xi1}--\ref{tab_taste_mass_ahisq}, \ref{tab_nomura} and \ref{tab_levkova_pisplit} were performed independently by two authors with different software: \texttt{lsqfit}~\cite{lsqfit} and the software similar to the one used in Ref.~\cite{MILC:2010pul}. The final values quoted in the tables are produced by the following procedure:
\begin{enumerate}
\item If necessary (for complicated cases such as heavier pion tastes), perform uncorrelated fits to find reasonable starting guesses for the fit parameters.
\item Perform correlated fits on the full ensemble to find where the ground state energy reaches a plateau and $\chi^2/\mbox{dof}$ is about 1.
\item Choose and keep constant $\tau_{min}$ consistent with point 2, and keep $\tau_{max}=N_\tau/2-1$.
\item Perform correlated fits (with the covariance matrix from the full ensemble) with fixed $\tau_{min}$, $\tau_{max}$ in each jackknife bin.
\item Propagate errors on the ground state energies (and the functions thereof such as pion splittings) with the jackknife procedure.
\end{enumerate}
Thus, the final quoted errors are propagated with jackknife and they are typically large enough to accommodate the systematic errors coming from varying $\tau_{min}$.

\subsection{Fermion anisotropy and strange quark mass tuning}
\label{sec_fermion_ani}

In our setup, the main parameter controlling the fermionic part of the action is the bare strange quark mass $a_\sigma m_s$. Its value is tuned such that the mass of the fictitious $\eta_{s\bar s}$ meson is approximately equal to $M_{s\bar s}^{phys}=685.8$~MeV, as computed by fitting lattice data to chiral perturbation theory in Ref.~\cite{Davies:2009tsa}. Similarly to the bare gauge anisotropy tuning, the ultimate goal is to determine the mapping $(a_\sigma m_s,\xi_0^f)\to(M_{s\bar s},\xi^f)$, where $\xi^f$ is the renormalized fermion anisotropy, that can be inverted and would allow one to predict what bare parameters $a_\sigma m_s$, $\xi_0^f$ are needed to produce an ensemble with desired properties. The fermion anisotropy may depend on the quark mass. To simplify tuning and define the action to have only one bare fermion anisotropy parameter independently of what quarks are included, we define the fermion anisotropy of an ensemble as the one evaluated at the strange quark mass that corresponds to the $\eta_{s\bar s}$ at the physical mass.
This choice is dictated by convenience, as it should not matter and different choices, \textit{e.g.}, $a_\sigma m_s/2$ or $a_\sigma m_s/27$ being the reference point, should give the same continuum limit. (And, as shown below, the dependence of the bare fermion anisotropy $\xi_0^f$ on the quark mass for aHISQ is flat across a large mass range.

To determine the renormalized fermion anisotropy we use the dispersion relation method. We performed tuning for both naive and highly improved staggered quarks to investigate the effect of smearing on the staggered spectrum and to also validate our code and tuning procedure by comparing with the naive staggered simulations in Refs.~\cite{Nomura:2004qu} and \cite{Levkova:2006gn}.

We discuss the parameter tuning of the naive staggered quark action first. At this stage we consider $(\beta,\xi_0)$ fixed and for brevity suppress the parameters that are fixed in the following. For illustration, we use $(\beta=6.87348, \xi_0=1.15792)$ ensemble listed in Table~\ref{tab_pg_ens} that has the renormalized anisotropy $\xi=1.2$.

\begin{SCfigure}
\includegraphics[width=0.5\textwidth]{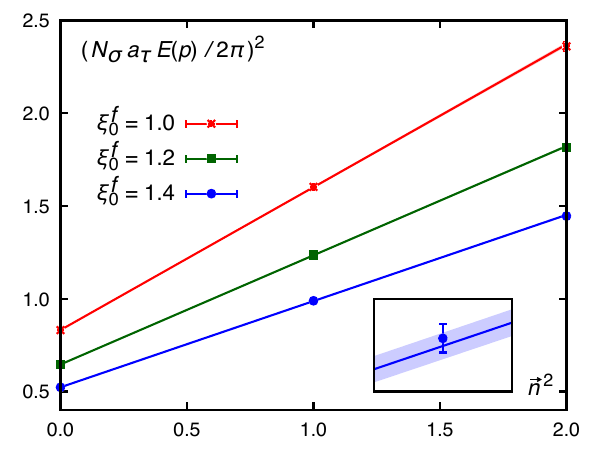}
\caption{Fits to the dispersion relation for naive staggered fermions on the ensemble $(\beta=6.87348, \xi_0=1.15792)$ that has the renormalized anisotropy $\xi=1.2$. The bare quark mass is $a_\sigma m_s=0.02$ and the bare fermion anisotropy is varied $\xi_0^f=1$, $1.2$ and $1.4$. The slopes of the fits produce, respectively, $\xi^f=1.13958(436)$, $\xi^f=1.30358(351)$ and $\xi^f=1.46703(384)$.\label{fig_disp_xi12_naive}}
\end{SCfigure}

In the first step of our fermion anisotropy tuning process we keep the bare strange quark mass $a_\sigma m_s$ fixed. We measure the Goldstone pseudoscalar meson correlation functions for the $\eta_{s\bar s}$ meson at the three lowest spatial momenta $a_\sigma \vec{p}=2\pi \vec{n}/N_\sigma$ for several bare fermion anisotropies $\xi_0^f$. In more convenient integer units they are $\vec{n}=(0,0,0)$, $\vec{n}=(1,0,0)$ and $\vec{n}=(1,1,0)$, and correspond to $(N_\sigma a_\sigma \vec{p}/2\pi)^2=\vec{n}^2=0,1,2$. 
The energy squared plotted in similarly rescaled units as $(N_\sigma a_\tau E(p)/2\pi)^2$ vs $\vec{n}^2$ for the $\xi=1.2$ ensemble at $a_\sigma m_s=0.02$ is shown in Fig.~\ref{fig_disp_xi12_naive} for three bare fermion anisotropies $\xi_0^f=1$, $1.2$ and $1.4$. By fitting the dispersion relation~(\ref{eq_disprel}) with a linear polynomial, we determine the renormalized anisotropy $\xi^f$ for a given bare $\xi_0^f$. In the units of Fig.~\ref{fig_disp_xi12_naive} the slope is simply $1/(\xi^f)^2$. The errors are propagated with the jackknife procedure. The errorbars are not visible on the scale of the figure and the lines, in fact, have error bands. This is illustrated in the inset of Fig.~\ref{fig_disp_xi12_naive} by magnifying the vicinity of the $\vec{n}^2=1$ point for $\xi_0^f=1.4$.

\begin{figure}
\includegraphics[width=0.49\textwidth]{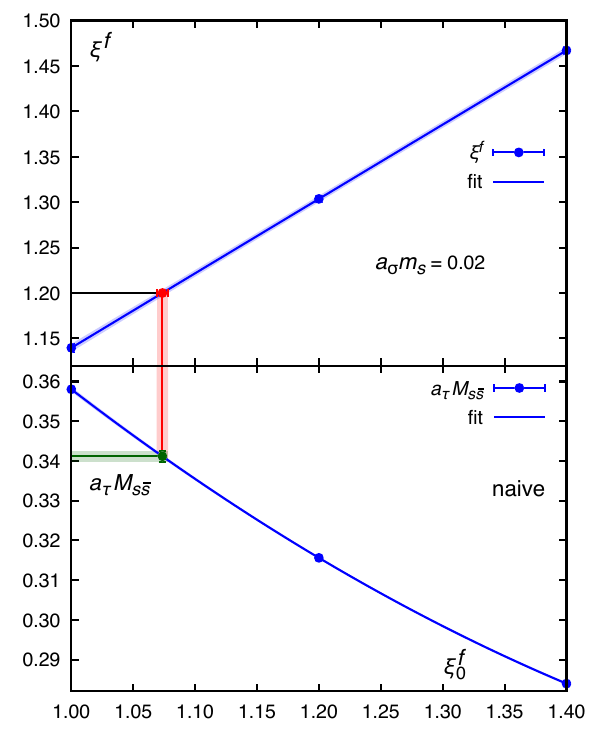}\hfill
\includegraphics[width=0.49\textwidth]{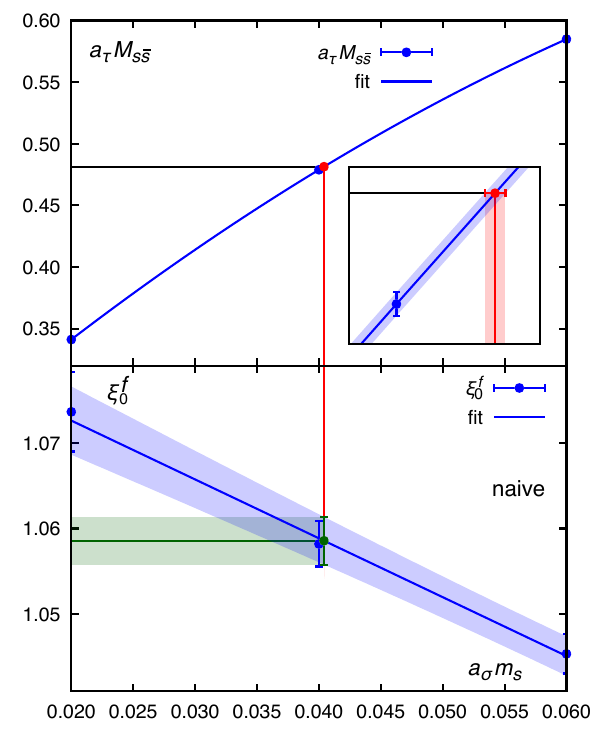}
\caption{(Left panel) Finding the bare fermion anisotropy $\xi_0^f$ that corresponds  to the renormalized fermion anisotropy $\xi_f=\xi=1.2$ (top) and then computing the $\eta_{s\bar s}$ at that $\xi_0^f$ (bottom) for naive staggered fermions at the strange quark mass $a_\sigma m_s=0.02$. (Right panel) Finding the bare strange quark mass at which the $\eta_{s\bar s}$ is at the physical value, $a_\sigma m_s=0.040412(44)$ (top) and computing the bare fermion anisotropy at that mass, $\xi_0^f=1.0586(28)$.\label{fig_xif_Mss_xi12_naive}}
\end{figure}

Once the renormalized fermion anisotropy $\xi^f$ and the $\eta_{s\bar s}$ mass are determined for each bare fermion anisotropy $\xi_0^f$, we can fit them as functions of $\xi_0^f$. We use a quadratic polynomial, however, given that there are only three bare anisotropies, this is simply an interpolation. The data together with the fits are shown in the left panel of Fig.~\ref{fig_xif_Mss_xi12_naive}. 
By solving $\xi^f(\xi_0^f)=\xi=1.2$ for $\xi_0^f$ (upper panel) we determine the bare fermion anisotropy $\xi_0^{f*}=1.074(46)$ that corresponds to the renormalized anisotropy $\xi^f=1.2$. Evaluating $a_\tau M_{s\bar s}(\xi_0^{f*})$ gives the $\eta_{s\bar s}$ mass, $a_\tau M_{s\bar s}=0.34117(43)$ that also corresponds to $\xi=1.2$. Now the bare fermion anisotropy is properly tuned and the renormalized anisotropy is the same for both, gauge and fermion parts of the action. However, the mass of the $\eta_{s\bar s}$ meson is at some arbitrary value, since we held $a_\sigma m_s=0.02$ fixed until now. Thus, at the next tuning stage, the process described above is repeated for several values of the bare strange quark mass, in this case $a_\sigma m_s=0.04$ and $0.06$. 
As a result, we get the dependence $\xi_0^f(a_\sigma m_s)$ for fixed $\xi^f=1.2$ (\textit{i.e.}, what bare anisotropy is needed to produce the renormalized fermion anisotropy at a given $a_\sigma m_s$) and the dependence $a_\tau M_{s\bar s}(a_\sigma m_s)$, also for fixed $\xi_f=1.2$. These two functions (with the data and the fits) are shown in the right panel of Fig.~\ref{fig_xif_Mss_xi12_naive}. The $a_\tau M_{s\bar s}(a_\sigma m_s)$ (top panel of the right panel of Fig.~\ref{fig_xif_Mss_xi12_naive}) function has some curvature and is fit (in fact, interpolated since there are only three points) with a quadratic polynomial. The $\xi_0^f(a_\sigma m_s)$ dependence is consistent with a linear fit. For the final tuning step we compute the mass of the fictitious $\eta_{s\bar s}$ meson in units of the temporal lattice spacing: $a_\tau M_{s\bar s}=a_\sigma/\xi\cdot M_{s\bar s}^{phys}=0.166218\mbox{ fm}/1.2\cdot 685.8\mbox{ MeV}/197.327\mbox{ MeV}\cdot\mbox{fm}\approx0.4814$. Solving $a_\tau M_{s\bar s}(a_\sigma m_s^*)=0.4814$ for $a_\sigma m_s^*$ (top panel) and then evaluating $\xi_0^f(a_\sigma m_s^*)$ (bottom panel) gives the final parameters: for naive staggered fermions on the quenched ensemble $(\beta=6.87348, \xi_0=1.15792)$ the pair $(a_\sigma m_s=0.040412(44),\xi_0^f=1.0586(28))$ is on the $\xi=1.2$ LCRA with the mass of the $\eta_{s\bar s}$ at the physical value.

Using the procedure described above, we tuned three anisotropic naive staggered ensembles with the renormalized anisotropy $\xi=1.2$, $1.5$ and $2$ at fixed spatial lattice spacing $a_\sigma\approx0.1665$~fm. One should note that, despite the fact that we only need to compute strange quark propagators, the tuning process is quite computationally demanding. In this case, very similar to tuning of the gauge anisotropy, we needed to generate a grid in the parameter space that spanned:
\begin{itemize}
\item $\xi=1.2$: $a_\sigma m_s=\{0.02,0.04,0.06\}\times \xi_0^f=\{1,1.2,1.4\}$,
\item $\xi=1.5$: $a_\sigma m_s=\{0.03,0.05,0.07,0.09\}\times \xi_0^f=\{1.05,1.25,1.45,1.65\}$,
\item $\xi=2\phantom{.0}$: $a_\sigma m_s=\{0.025,0.035,0.045\}\times \xi_0^f=\{1.2,1.6,2\}$.
\end{itemize}
For each point on the $(a_\sigma m_s,\xi_0^f)$ grid we need to compute correlation functions at three momenta.

\begin{figure}
\includegraphics[width=0.49\textwidth]{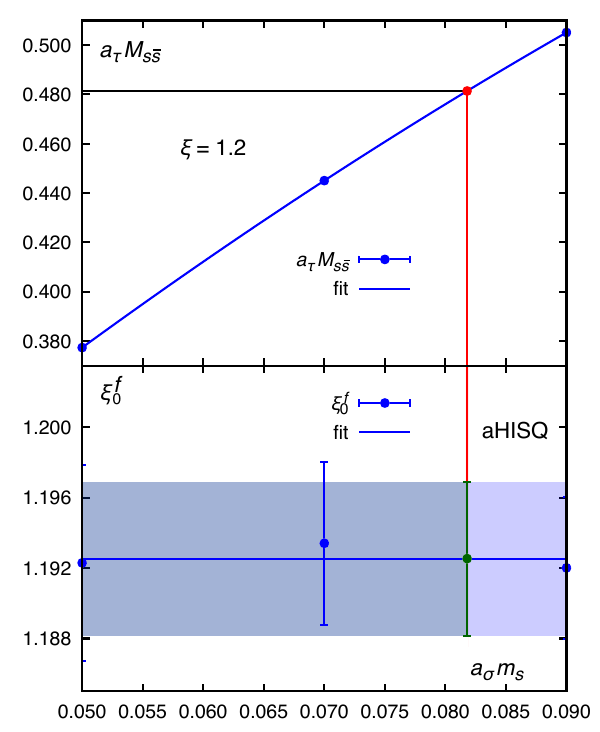}\hfill
\includegraphics[width=0.49\textwidth]{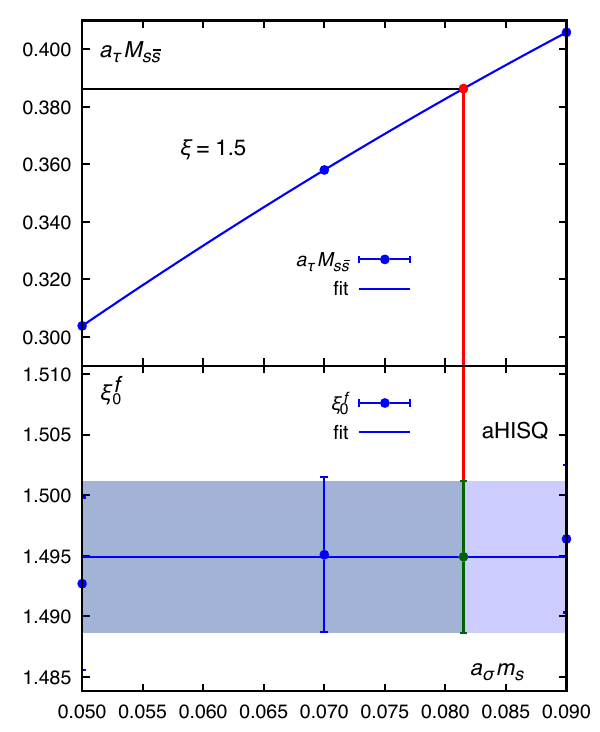}
\caption{Final tuning steps that determine the strange quark mass that corresponds to the $\eta_{s\bar s}$ at the physical point (upper panels) and the corresponding bare fermion anisotropy (lower panels) for the aHISQ action on the ensembles with the renormalized anisotropy $\xi=1.2$ and $\xi=1.5$. These are the same tuning steps as shown in the right panel of Fig.~\ref{fig_xif_Mss_xi12_naive} for the naive staggered action.\label{fig_xif_Mss_xi12_15_ahisq}}
\end{figure}

We followed the same tuning protocol for the aHISQ action at two anisotropies $\xi=1.2$ and $\xi=1.5$. The final stages of the tuning process are shown in the left and right panels of Fig.~\ref{fig_xif_Mss_xi12_15_ahisq}, respectively. In that case the parameter grid spanned:
\begin{itemize}
\item $\xi=1.2$: $a_\sigma m_s=\{0.05,0.07,0.09\}\times \xi_0^f=\{1,1.2,1.4\}$,
\item $\xi=1.5$: $a_\sigma m_s=\{0.05,0.07,0.09\}\times \xi_0^f=\{1.05,1.25,1.45,1.65\}$.
\end{itemize}
While the functional dependence $a_\tau M_{s\bar s}(a_\sigma m_s)$ is similar between the naive staggered and aHISQ actions, the striking feature of aHISQ is almost no dependence of the bare fermion anisotropy on the quark mass $\xi_0^f(a_\sigma m_s)$: in the range where the $\eta_{s\bar s}$ mass changes by about 25\%, the variation in the bare fermion anisotropy is about 0.7\% and can be fit with a constant, Fig.~\ref{fig_xif_Mss_xi12_15_ahisq}. Moreover, the value of the tuned bare fermion anisotropy for aHISQ is within 0.6\% ($1.7\sigma$) for $\xi=1.2$ and 0.3\% ($0.8\sigma$) for $\xi=1.5$ of the target renormalized anisotropy.
(For $\xi=1.5$ we also have measurements at $a_\sigma m_s=0.03$, as indicated in Table~\ref{tab_tune_ahisq_xi15}. We do not include them in the tuning process, shown in Fig.~\ref{fig_xif_Mss_xi12_15_ahisq}, as we are interested in the strange quark mass not too far from the physical value. However, even if that point is included in the right panel of Fig.~\ref{fig_xif_Mss_xi12_15_ahisq}, the fit for $\xi_0^f$ (lower panel) is still consistent with a constant.)

The flatness of the $\xi_0^f(a_\sigma m_s)$ dependence means that tuning of aHISQ can be drastically simplified. If the trend $\xi_0^f\approx\xi$ persists to larger anisotropies, one can simply set $\xi_0^f=\xi$ (target renormalized anisotropy) in the fermion part of the action and only tune the strange quark mass. To check that, however, we adopted a simplified tuning procedure for aHISQ that still allows us to tune both the bare fermion anisotropy and strange quark mass. Namely, instead of measuring the dispersion relation on the full grid of masses and bare anisotropies, we first fix the strange quark mass close to the anticipated target value, and vary the bare fermion anisotropy only at that mass. From the fit we find proper tuned value of $\xi_0^f$ (that gives $\xi^f=\xi$), and then vary the strange quark mass only at that bare anisotropy. In other words, for aHISQ one can reduce two-dimensional grid search to two one-dimensional searches, where only one of them requires non-zero momentum measurements.

With the simplified procedure described above we tuned aHISQ action at the additional renormalized anisotropies $\xi=1.1$, $2$, $4$ and $8$.
The values of the ground state energies for aHISQ used in the tuning process are documented in Tables~\ref{tab_tune_ahisq_xi1}--\ref{tab_tune_ahisq_xi8} and the predicted $a_\sigma m_s$ and $\xi_0^f$ for various ensembles in Table~\ref{tab_splittings_run_params}. It has to be noted that in those tables we documented the values in units of the temporal lattice spacing, $a_\tau E$, directly as they are determined from the fits. For the discussion in the next section however, in order to be able to compare different anisotropies, we need to convert everything to common units. A convenient choice is $w_0$ units: $w_0E=(w_0/a_\sigma)\xi(a_\tau E)$. In other words, the $a_\tau E$ values from the tables are multiplied by the gradient flow scale $w_0/a_\sigma$, listed in the seventh column of Table \ref{tab_pg_ens}, and by the renormalized anisotropy $\xi$, listed in the third column of the same table.

\subsection{Pion taste masses and splittings}
\label{sec_pion_split}

\begin{figure}
\includegraphics[width=0.99\textwidth]{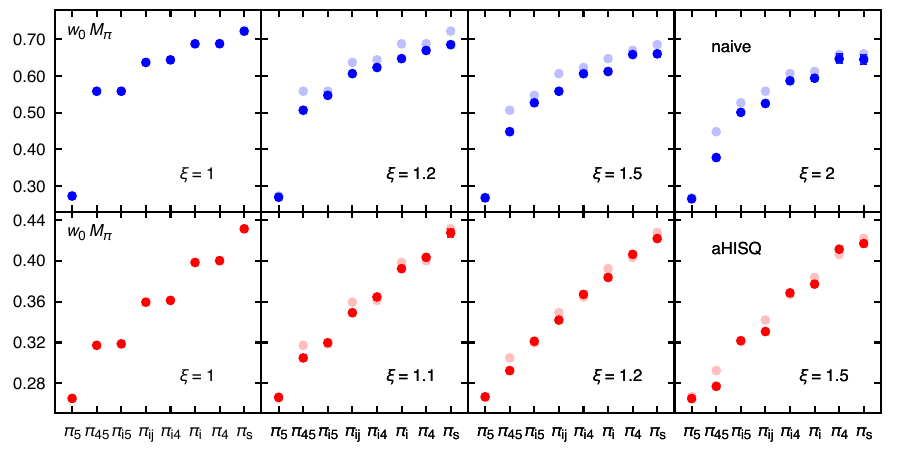}
\caption{Progression of pion taste masses, plotted as dimensionless combinations $w_0M_\pi$, from isotropic $\xi=1$ case to anisotropy up to 2 for naive (upper panel) and 1.5 for aHISQ (lower panel) action. The Goldstone pion $\pi_5$ is tuned to approximately same value for all anisotropies. Pale points represent the values of the masses at the anisotropy on the plot immediately to the left.\label{fig_pion_mass}}
\end{figure}

Our main focus is to understand how to tune the aHISQ action for future dynamical simulations. We also need to study the pattern of the pion taste multiplets, as taste-breaking effects are often the dominant discretization effects with staggered actions. In this initial phase, we study the dependence of the pion taste masses on the anisotropy at fixed spatial lattice spacing.
The bare light quark mass is set to $a_\sigma m_l=a_\sigma m_s/5$.
For lower anisotropies, we also measured the pion taste masses with the naive staggered action. Those allow us to check correctness of our workflow (from ensemble generation to the final fits), and also compare two different actions, where we, in fact, observe some qualitative differences. To illustrate the effect, we plot the pion taste masses at different anisotropies in Fig.~\ref{fig_pion_mass}. The top panel displays the values for the naive action and the bottom for the aHISQ action. The pion mass is in units of $1/w_0$. The anisotropy increases from left to right, and since the dependence on anisotropy at low values is much steeper for aHISQ, we illustrate the change for the naive staggered action on $\xi=1$, $1.2$, $1.5$ and $2$, and for aHISQ on $\xi=1$, $1.1$, $1.2$ and $1.5$ ensembles. At the anisotropies $\xi>1$ the pale points represent the values at the previous anisotropy. From the progressions in Fig.~\ref{fig_pion_mass} one can observe the following: for the naive staggered action all heavier pion tastes move down, as the anisotropy increases, approaching the new degeneracy pattern expected for $\xi\gg1$, as discussed in Sec.~\ref{sec_ani_stag_spec}.
For aHISQ, the $\xi_4\xi_5$, $\xi_i\xi_j$, $\xi_i$ and $1$ tastes move down, while the $\xi_i\xi_5$,  $\xi_i\xi_4$ and $\xi_4$ move up at much slower rate, compared to the first four. (On the scale of the figure the $\xi_i\xi_5$ taste appears constant, but it also moves up, as will become apparent shortly.) For the ease of notation, we use two ways to denote pion tastes: with the explicit taste matrix-structure, \textit{e.g.}, $\xi_i\xi_5$, and as $\pi$ with a proper index, \textit{e.g.}, $\pi_{i5}$. The taste structure $1$ is denoted $\pi_s$.

Now we consider the quadratic splittings from the lowest, Goldstone pion mass:
\begin{equation}
\delta M_\pi\equiv M_\pi^2-M_{\pi_5}^2,
\end{equation}
where the index $\pi$ denotes all taste structures except of $\xi_5$. For the ease of comparison between the naive staggered and aHISQ, we normalize by the splittings for the heaviest taste, $\pi_s$, at the anisotropy $\xi=1$:
\begin{equation}
\label{eq_Mpi_norm_pis}
\Delta_\pi\equiv\frac{\delta M_\pi}{\left.\delta M_{\pi_s}\right|_{\xi=1}}.
\end{equation}

\begin{figure}
\includegraphics[width=0.98\textwidth]{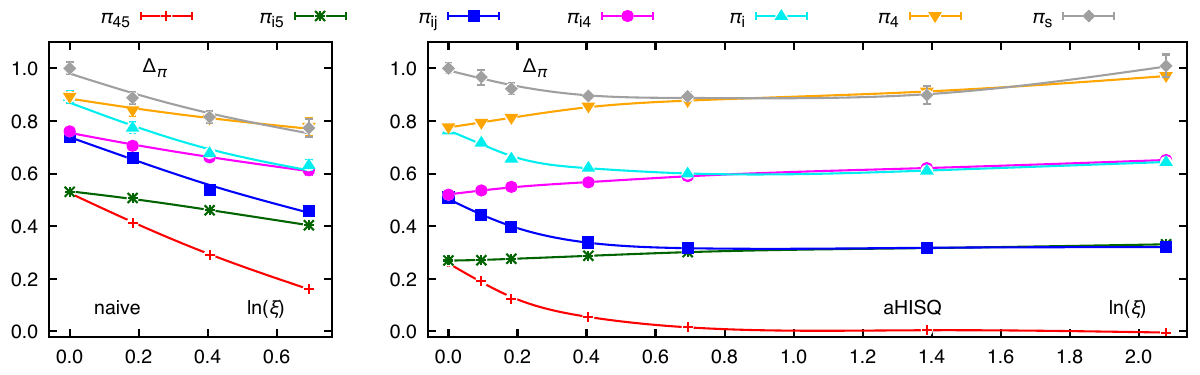}
\caption{Splittings of pion taste multiplets from the Goldstone pion, normalized by the value for the singlet taste at $\xi=1$, Eq.~(\ref{eq_Mpi_norm_pis}), for the naive staggered (left) and aHISQ (right) action, as function of $\ln(\xi)$. \label{fig_pion_split}}
\end{figure}

The normalized splittings $\Delta_\pi$ for the naive staggered and aHISQ action are shown in left and right panels of Fig.~\ref{fig_pion_split}, respectively.
The lines are not fits but simple spline interpolations to guide the eye.
Apart from showing the obvious benefits of the smeared actions,--\textit{i.e.}, not only the naive taste masses are larger in magnitude, Fig.~\ref{fig_pion_mass}, the relative gap between the Goldstone and the next taste is twice larger for the naive staggered action compared to HISQ,--Fig.~\ref{fig_pion_split} highlights the qualitatively different behavior in the multiplet formation when anisotropy is increased. The naive taste splittings decrease, and this happens at a different but comparable rate to allow for the new pattern formation which is almost complete at the largest considered anisotropy $\xi=2$ (except of the $\xi_4\xi_5$ not joining the $\xi_5$ taste yet). For aHISQ at low anisotropy, all odd pion taste splittings (using the ordering in the legend of Fig.~\ref{fig_pion_split}) decrease at fast rate, while all the even ones increase, at a much slower rate. Moreover, the increase of the even splittings appears to become the dominant effect at higher anisotropy, and the odd splittings (except of the $\xi_4\xi_5$ that must become degenerate with $\xi_5$) must follow to maintain the new degeneracy pattern. For that reason, at some anisotropy between $2$ and $4$ the odd splittings acquire a minimum and start slowly increasing, following the even splittings. The effect is mild but resolvable within the errors.

\begin{figure}
\includegraphics[width=0.98\textwidth]{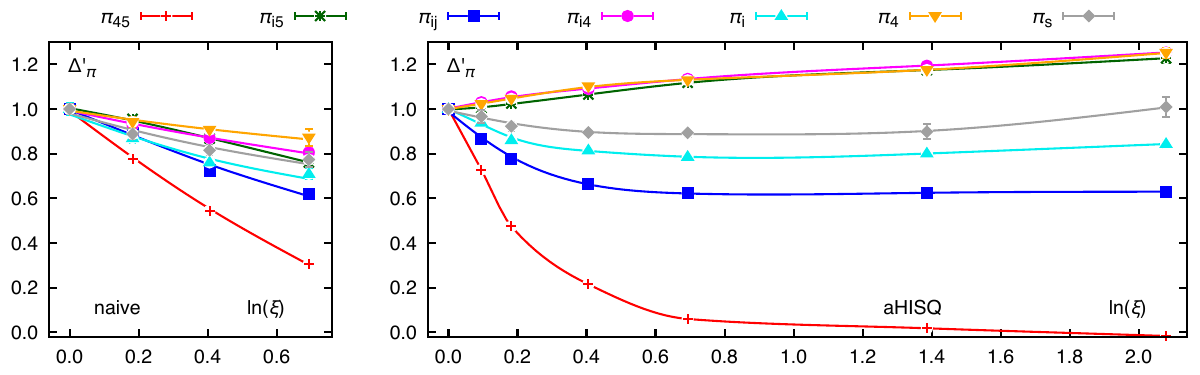}
\caption{Splittings of pion taste multiplets from the Goldstone pion, normalized by their individual values at $\xi=1$, Eq.~(\ref{eq_Mpi_norm_xi1}), for the naive staggered (left) and aHISQ (right) action, as function of $\ln(\xi)$.\label{fig_pion_split2}}
\end{figure}

Another way to look at the taste splittings is to normalize them by their individual corresponding value at the renormalized anisotropy $\xi=1$:
\begin{equation}
\label{eq_Mpi_norm_xi1}
\Delta'_\pi\equiv\frac{\delta M_\pi}{\left.\delta M_{\pi}\right|_{\xi=1}}.
\end{equation}
In that normalization, they are plotted in Fig.~\ref{fig_pion_split2}. The lines are not fits but spline interpolations to guide the eye. While for the naive staggered action there is no specific pattern, we observe that for aHISQ the even splittings that slightly increase ($\xi_i\xi_5$, $\xi_i\xi_4$ and $\xi_4$) collapse onto an almost universal curve. In other words, it may be that at the leading order their dependence on anisotropy is the same, and they only differ by multiplicative factors.

The observed patterns for the aHISQ pion taste splittings prompted our attempt to develop a qualitative model, described in the next section, that describes the splittings dependence on anisotropy. 

\subsection{Staggered chiral perturbation theory and a model for pion taste mass splittings}
\label{sec_schipt}

\begin{figure}
\includegraphics[width=0.98\textwidth]{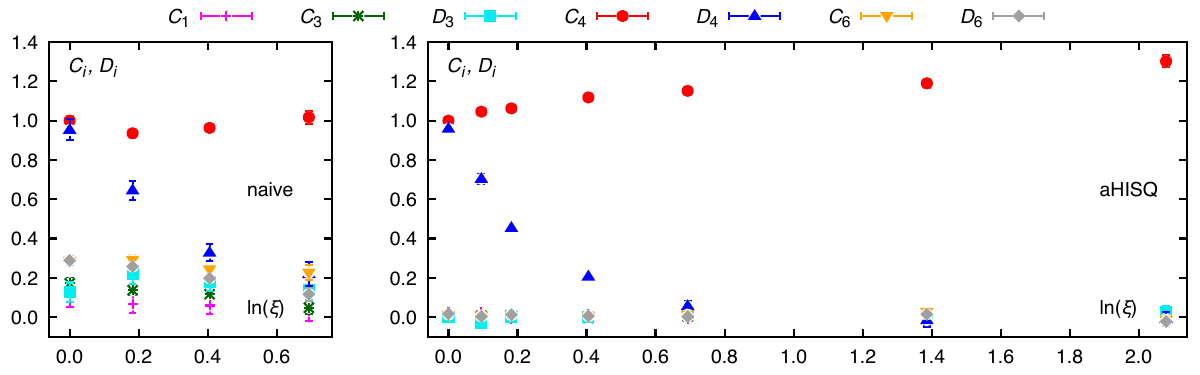}
\caption{The low-energy constants $C_i$ and $D_i$ as functions of $\ln(\xi)$ for the naive staggered (left) and aHISQ (right) action. For comparison, the values on the left are divided by $C_4(\xi=1)=0.0974(22)$ for the naive staggered and on the right by $C_4(\xi=1)=0.02914(37)$ for the aHISQ action.
\label{fig_chipt_lec}
}
\end{figure}

The dimension-6 operators contributing to the pion taste mass splittings have been identified in Refs.~\cite{Luo:1997tt,Lee:1999zxa}. For the isotropic case the chiral perturbation theory Lagrangian has the following form:
\begin{equation}
\mathcal{L}(\Sigma)=\frac{a^4f^2}{8}\text{Tr}\Big(\partial_\mu \Sigma \partial_\mu \Sigma^\dagger\Big)-\frac{a^4}{4}Bf^2\text{Tr}\Big(\mathcal{M}\Sigma+\mathcal{M}\Sigma^\dagger\Big)+\frac{a^4m_0^2}{24}\big(\text{Tr}\Phi\big)^2+\mathcal{V},
\end{equation}
where the field $\Sigma$ is defined as $\Sigma=\exp(i\Phi/f)$, with $f,B$ low energy constants and the term before the last term being relevant only for the taste- and flavor-singlet $\eta'$~\cite{Bernard:2001yj}. The dimension-6 potential $\mathcal{V}$ contains the operators responsible for taste splittings, and we deliberately do not factor out $a^6$, as will become clear in a moment. It has two parts $\mathcal{V}=\mathcal{U}+\mathcal{U}'$, where $\mathcal{U}'$ affects only flavor-singlet states, as explained in Ref.~\cite{Aubin:2003mg}, and we thus focus on the $\mathcal{U}$ in the following:
\begin{eqnarray}
-\mathcal{U}
&=& a^6\sum_k C_k \mathcal{O}_k \nonumber\\
&=& a^6C_1\text{Tr}\big(\xi_{5}
\Sigma\xi_{5}\Sigma^\dagger\big)
\nonumber\\
&+&a^6C_3 \frac{1}{2}\sum_\nu
\Bigl[
\text{Tr}\big(\xi_{\nu}\Sigma
\xi_{\nu}\Sigma\big)
+ \text{Tr}\big(\xi_{\nu}\Sigma^\dagger
\xi_{\nu}\Sigma^\dagger\big)
\Bigr]
\nonumber\\
&+& a^6C_4 \frac{1}{2}\sum_\nu
\Bigl[
\text{Tr}\big(\xi_{\nu5}\Sigma
\xi_{5\nu}\Sigma\big)
+\text{Tr}\big(\xi_{\nu5}\Sigma^\dagger
\xi_{5\nu}\Sigma^\dagger\big)
\Bigr]
\nonumber\\
&+& a^6C_6 \sum_{\mu<\nu}
\text{Tr}\big(
\xi_{\mu\nu}\Sigma
\xi_{\nu\mu}\Sigma^\dagger
\big).\label{eq_chi_pot_U}
\end{eqnarray}
Anisotropy appears on the level of the fundamental theory and its effect on the level of the effective theory is not obvious. A proper procedure requires extending the Lee-Sharpe analysis~\cite{Lee:1999zxa} to anisotropic case. This is beyond the scope of the present paper, and is a possible direction for a future study. Here, we would like to find an empirical modification of the staggered chiral perturbation theory Lagrangian that can describe the anisotropy dependence observed in the data.
The simplest way to transition to the anisotropic case is then to introduce appropriate factors of the temporal lattice spacing for the operators that contain temporal direction. Proceeding as in the isotropic case~\cite{Lee:1999zxa} by expanding the $\Sigma$ field to second order, one arrives at the following pattern for the taste splittings:
\begin{eqnarray}
\delta M_{45}(\xi) &=&
\frac{16a_\sigma^2}{f^2}
\left(
C_1 + 3C_3 + \frac{1}{\xi^2}C_4 + 3\frac{1}{\xi^2}C_6
\right)\label{eq_anichipt_split_noDs_first}
\\
\delta M_{i5}(\xi) &=&
\frac{16a_\sigma^2}{f^2}
\left(
C_1 + \left(2+\frac{1}{\xi^2}\right)C_3 + C_4
+ \left(2+\frac{1}{\xi^2}\right)C_6
\right)
\\
\delta M_{ij}(\xi) &=&
\frac{16a_\sigma^2}{f^2}
\left(1+\frac{1}{\xi^2}\right)
\left(
C_3 + C_4 + 2C_6
\right)
\\
\delta M_{i4}(\xi) &=&
\frac{16a_\sigma^2}{f^2}
\,2\left(
C_3 + C_4 + \left(1+\frac{1}{\xi^2}\right)C_6
\right)
\\
\delta M_i(\xi) &=&
\frac{16a_\sigma^2}{f^2}
\left(
C_1 + C_3 + \left(2+\frac{1}{\xi^2}\right)C_4
+ \left(2+\frac{1}{\xi^2}\right)C_6
\right)
\\
\delta M_4(\xi) &=&
\frac{16a_\sigma^2}{f^2}
\left(
C_1 + \frac{1}{\xi^2}C_3 + 3C_4
+ 3\frac{1}{\xi^2}C_6
\right)
\\
\delta M_s(\xi) &=&
\frac{16a_\sigma^2}{f^2}
\left(
\left(3+\frac{1}{\xi^2}\right)C_3
+\left(3+\frac{1}{\xi^2}\right)C_4
\right)\label{eq_anichipt_split_noDs_last}
\end{eqnarray}
It is known~\cite{Lee:1999zxa} that for smeared actions, such as HISQ, the splittings are dominated by the $C_4$ coefficient with the rest being close to zero. (We illustrate this with the actual data later.) We find, however, that the form in Eqs.~(\ref{eq_anichipt_split_noDs_first})--(\ref{eq_anichipt_split_noDs_last}) is inconsistent with the data. It predicts that for aHISQ the masses of the $\xi_i\xi_5$, $\xi_i\xi_4$ and $\xi_4$ tastes stay constant, while the $\xi_4\xi_5$, $\xi_i\xi_j$, $\xi_i$ and $1$ tastes approach the partner values as $1/\xi^2$, however, Fig.~\ref{fig_pion_split2} clearly shows that $\xi_i\xi_5$, $\xi_i\xi_4$ and $\xi_4$ splittings increase with anisotropy. Moreover, for the naive staggered action the $\delta M_{45}$ splitting would approach $C_1+3C_3$ rather than 0, when the $\xi_4\xi_5$ taste is expected to become degenerate with $\xi_5$. Thus, one needs to introduce anisotropy dependence in the low-energy constants. This alone, however, does not allow us to describe the data. We therefore introduce a further modification of the potential in Eq.~(\ref{eq_chi_pot_U}). We assume that for anisotropic case the low-energy constants for the temporal components are different from the spatial ones. The potential then takes the following form:
\begin{eqnarray}
-\mathcal{U}_{ani}
&=& a_\sigma^6C_1(\xi)\text{Tr}\big(\xi_{5}
\Sigma\xi_{5}\Sigma^\dagger\big)
\nonumber\\
&+&a_\sigma^6C_3(\xi) \frac{1}{2}\sum_j
\Bigl[
\text{Tr}\big(\xi_j\Sigma
\xi_j\Sigma\big)
+ \text{Tr}\big(\xi_j\Sigma^\dagger
\xi_j\Sigma^\dagger\big)
\Bigr]\nonumber\\
&+&a_\sigma^4a_\tau^2C'_3(\xi) \frac{1}{2}\Bigl[
\text{Tr}\big(\xi_{4}\Sigma
\xi_{4}\Sigma\big)
+ \text{Tr}\big(\xi_4\Sigma^\dagger
\xi_4\Sigma^\dagger\big)
\Bigr]
\nonumber\\
&+& a_\sigma^6C_4(\xi) \frac{1}{2}\sum_j
\Bigl[
\text{Tr}\big(\xi_{j5}\Sigma
\xi_{5j}\Sigma\big)
+\text{Tr}\big(\xi_{j5}\Sigma^\dagger
\xi_{5j}\Sigma^\dagger\big)
\Bigr]\nonumber\\
&+&a_\sigma^4a_\tau^2C'_4(\xi) \frac{1}{2}
\Bigl[
\text{Tr}\big(\xi_{45}\Sigma
\xi_{54}\Sigma\big)+\text{Tr}\big(\xi_{45}\Sigma^\dagger
\xi_{54}\Sigma^\dagger\big)
\Bigr]
\nonumber\\
&+& a_\sigma^6C_6(\xi) \sum_{j<k}
\text{Tr}\big(
\xi_{jk}\Sigma
\xi_{kj}\Sigma^\dagger
\big)\nonumber\\
&+&a_\sigma^4a_\tau^2C'_6(\xi) \sum_j
\text{Tr}\big(
\xi_{j4}\Sigma
\xi_{4j}\Sigma^\dagger
\big).
\end{eqnarray}
Factoring out the spatial lattice spacing $a_\sigma^6$ produces explicit factors of anisotropy $1/\xi^2$ that always accompany the $C'_i(\xi)$ low-energy constants, so for convenience we replace them with $D_i(\xi)\equiv C'_i(\xi)/\xi^2$.
All terms then share a common factor $a_\sigma^6$, and the final potential that we use, rescaled by that factor, is then:
\begin{eqnarray}
-\mathcal{U}_{ani}/a_\sigma^6
&=& C_1(\xi)\text{Tr}\big(\xi_{5}
\Sigma\xi_{5}\Sigma^\dagger\big)
\nonumber\\
&+&C_3(\xi) \frac{1}{2}\sum_j
\Bigl[
\text{Tr}\big(\xi_j\Sigma
\xi_j\Sigma\big)
+ \text{Tr}\big(\xi_j\Sigma^\dagger
\xi_j\Sigma^\dagger\big)
\Bigr]\nonumber\\
&+&D_3(\xi) \frac{1}{2}\Bigl[
\text{Tr}\big(\xi_{4}\Sigma
\xi_{4}\Sigma\big)
+ \text{Tr}\big(\xi_4\Sigma^\dagger
\xi_4\Sigma^\dagger\big)
\Bigr]
\nonumber\\
&+&C_4(\xi) \frac{1}{2}\sum_j
\Bigl[
\text{Tr}\big(\xi_{j5}\Sigma
\xi_{5j}\Sigma\big)
+\text{Tr}\big(\xi_{j5}\Sigma^\dagger
\xi_{5j}\Sigma^\dagger\big)
\Bigr]\nonumber\\
&+&D_4(\xi) \frac{1}{2}
\Bigl[
\text{Tr}\big(\xi_{45}\Sigma
\xi_{54}\Sigma\big)+\text{Tr}\big(\xi_{45}\Sigma^\dagger
\xi_{54}\Sigma^\dagger\big)
\Bigr]
\nonumber\\
&+&C_6(\xi) \sum_{j<k}
\text{Tr}\big(
\xi_{jk}\Sigma
\xi_{kj}\Sigma^\dagger
\big)\nonumber\\
&+&D_6(\xi) \sum_j
\text{Tr}\big(
\xi_{j4}\Sigma
\xi_{4j}\Sigma^\dagger
\big).\label{eq_chipt_anipotential_final}
\end{eqnarray}
In general, the low-energy constants are functions of two lattice spacings, $a_\sigma$ and $a_\tau$. We remind the reader that in this study $a_\sigma={\rm const}$, thus the dependence is on a single parameter, $\xi$.
From this point on, for ease of notation, we include the constant prefactor $16a_\sigma^2/f^2$ into the definition of the low-energy constants, \textit{i.e.}, we rescale, \textit{e.g.}, $16a_\sigma^2/f^2C_i\to C_i$.
By definition, $D_i(\xi=1)\equiv C_i(\xi=1)$.
The taste splittings derived in the same way as before for the potential in Eq.~(\ref{eq_chipt_anipotential_final}) are:
\begin{IEEEeqnarray}{l l c r c r r c r r c r r c r r l r r c r r l}
\delta M_{45}&(\xi) &{}={}& C_1(\xi) &{}+{}& 3 & C_3(\xi) &&&&&&&{}+{}&& D_4(\xi) &&&&{}+{}&3& D_6(\xi)&,
\label{eq_anichipt_split_noDs_first_2}  \\
\delta M_{i5}&(\xi) &{}={}& C_1(\xi) &{}+{}& 2 & C_3(\xi) &{}+{}&& D_3(\xi) &{}+{}& & C_4(\xi) &&&&{}+{}& 2 & C_6(\xi) &{}+{}&& D_6(\xi)&,  \\
\delta M_{ij}&(\xi) &{}={}&&&& C_3(\xi) &{}+{}&& D_3(\xi) &{}+{}& & C_4(\xi) &{}+{}&&D_4(\xi)&{}+{}& 2 & C_6(\xi) &{}+{}&2& D_6(\xi)&,  \\
\delta M_{i4}&(\xi) &{}={}&&&2& C_3(\xi) &&&&{}+{}&2& C_4(\xi) &&&&{}+{}& 2 & C_6(\xi) &{}+{}&2& D_6(\xi)&,  \\
\delta M_{i}&(\xi) &{}={}& C_1(\xi) &{}+{}&& C_3(\xi) &&&&{}+{}&2& C_4(\xi) &{}+{}&&D_4(\xi)&{}+{}& 2 & C_6(\xi) &{}+{}&& D_6(\xi)&,  \\
\delta M_{4}&(\xi) &{}={}& C_1(\xi) &&&&{}+{}&& D_3(\xi) &{}+{}&3& C_4(\xi) &&&&&&&{}+{}&3& D_6(\xi)&,  \\
\delta M_{s}&(\xi) &{}={}&&& 3 & C_3(\xi) &{}+{}&& D_3(\xi) &{}+{}& 3 & C_4(\xi) &{}+{}&&D_4(\xi)&.&&&&&&\label{eq_anichipt_split_noDs_last_2}
\end{IEEEeqnarray}
Now there are seven taste splittings that depend on seven low-energy constants. Inverting the dependence allows us to express the low-energy constants in terms of the data for the taste splittings as functions of anisotropy:
\begin{IEEEeqnarray}{l l c c r r c r r c r r c r r c r r c r r c r r l}
C_1&(\xi) &{}={}& \frac{1}{8}( &&\delta M_{45}(\xi)&{}+{}&3&\delta M_{i5}(\xi)&{}-{}&3&\delta M_{ij}(\xi)&{}-{}&3&\delta M_{i4}(\xi)&{}+{}&3&\delta M_{i}(\xi)&{}+{}&&\delta M_{4}(\xi)&{}-{}&&\delta M_{s}(\xi)&),\\
C_3&(\xi) &{}={}& \frac{1}{8}( &&\delta M_{45}(\xi)&{}+{}&&\delta M_{i5}(\xi)&{}-{}&&\delta M_{ij}(\xi)&{}+{}&&\delta M_{i4}(\xi)&{}-{}&&\delta M_{i}(\xi)&{}-{}&&\delta M_{4}(\xi)&{}+{}&&\delta M_{s}(\xi)&),\\
D_3&(\xi) &{}={}& \frac{1}{8}( &-&\delta M_{45}(\xi)&{}+{}&3&\delta M_{i5}(\xi)&{}+{}&3&\delta M_{ij}(\xi)&{}-{}&3&\delta M_{i4}(\xi)&{}-{}&3&\delta M_{i}(\xi)&{}+{}&&\delta M_{4}(\xi)&{}+{}&&\delta M_{s}(\xi)&),\\
C_4&(\xi) &{}={}& \frac{1}{8}( &-&\delta M_{45}(\xi)&{}-{}&&\delta M_{i5}(\xi)&{}-{}&&\delta M_{ij}(\xi)&{}+{}&&\delta M_{i4}(\xi)&{}+{}&&\delta M_{i}(\xi)&{}+{}&&\delta M_{4}(\xi)&{}+{}&&\delta M_{s}(\xi)&),\\
D_4&(\xi) &{}={}& \frac{1}{8}( &&\delta M_{45}(\xi)&{}-{}&3&\delta M_{i5}(\xi)&{}+{}&3&\delta M_{ij}(\xi)&{}-{}&3&\delta M_{i4}(\xi)&{}+{}&3&\delta M_{i}(\xi)&{}-{}&&\delta M_{4}(\xi)&{}+{}&&\delta M_{s}(\xi)&),\\
C_6&(\xi) &{}={}& \frac{1}{8}( &-&\delta M_{45}(\xi)&{}+{}&&\delta M_{i5}(\xi)&{}+{}&&\delta M_{ij}(\xi)&{}+{}&&\delta M_{i4}(\xi)&{}+{}&&\delta M_{i}(\xi)&{}-{}&&\delta M_{4}(\xi)&{}-{}&&\delta M_{s}(\xi)&),\\
D_6&(\xi) &{}={}& \frac{1}{8}( &&\delta M_{45}(\xi)&{}-{}&&\delta M_{i5}(\xi)&{}+{}&&\delta M_{ij}(\xi)&{}+{}&&\delta M_{i4}(\xi)&{}-{}&&\delta M_{i}(\xi)&{}+{}&&\delta M_{4}(\xi)&{}-{}&&\delta M_{s}(\xi)&).
\end{IEEEeqnarray}
These dependencies are shown in Fig.~\ref{fig_chipt_lec}. As expected, for aHISQ (right panel) only $C_4(\xi)$ and $D_4(\xi)$ contribute to the taste splittings, while the other low-energy constants stay close to 0 also for $\xi>1$. For the naive staggered action (left panel) the situation is different: while the splittings are also dominated by $C_4(\xi)$ and $D_4(\xi)$, the other low-energy constants are also sizeable. It appears that $C_4(\xi)$ increases, similar to aHISQ. Importantly, $C_1(\xi)$ and $C_3(\xi)$ decrease with anisotropy, consistent with the expectation of $\delta M_{45}$ going to 0.

For aHISQ taste splittings the dependence on anisotropy is then simplified:
\begin{eqnarray}
\delta M_{45}(\xi) &=&
0\cdot C_4(\xi)+1\cdot D_4(\xi)
,
\label{eq_anichipt_split_noDs_first_hisq}
\\
\delta M_{i5}(\xi) &=&
1\cdot C_4(\xi)+0\cdot D_4(\xi)
,
\\
\delta M_{ij}(\xi) &=&
1\cdot C_4(\xi)+1\cdot D_4(\xi)
,
\\
\delta M_{i4}(\xi) &=&
2\cdot C_4(\xi)+0\cdot D_4(\xi)
,
\\
\delta M_i(\xi) &=&
2\cdot C_4(\xi)+1\cdot D_4(\xi)
,
\label{eq_anichipt_vec}
\\
\delta M_4(\xi) &=&
3\cdot C_4(\xi)+0\cdot D_4(\xi)
,
\\
\delta M_s(\xi) &=&
3\cdot C_4(\xi)+1\cdot D_4(\xi)
.
\label{eq_anichipt_split_noDs_last_hisq}
\end{eqnarray}
Equations~(\ref{eq_anichipt_split_noDs_first_hisq})--(\ref{eq_anichipt_split_noDs_last_hisq}) reveal an interesting feature of the aHISQ taste splittings. Consider, for instance, the vector, $\gamma_5\otimes\xi_i$ taste, Eq.~(\ref{eq_anichipt_vec}). Given that $\gamma_5=\gamma_1\gamma_2\gamma_3\gamma_4$, an interpolating operator requires 4 shifts to build the spin and 1 shift to build the taste structure. The spatial taste shift can balance one spatial spin shift, so in total there are $l_\sigma=2$ and $l_\tau=1$ unbalanced shifts, \textit{i.e.}, gauge links in the interpolating operator for the $\gamma_5\otimes\xi_i$ taste, as discussed in Sec.~\ref{sec_ani_stag_spec}. (From this consideration, the Goldstone taste, $\gamma_5\otimes\xi_5$ is, of course, local, as there are four spin and four taste shifts that can balance each other.) 
The values $l_\sigma=2$ and $l_\tau=1$ are precisely the coefficients in front of $C_4(\xi)$ and $D_4(\xi)$ in Eq.~(\ref{eq_anichipt_vec}). One can easily check that this interpretation of the coefficients applies to all aHISQ splittings in Eqs.~(\ref{eq_anichipt_split_noDs_first_hisq})--(\ref{eq_anichipt_split_noDs_last_hisq}).

\begin{figure}
\includegraphics[width=0.49\textwidth]{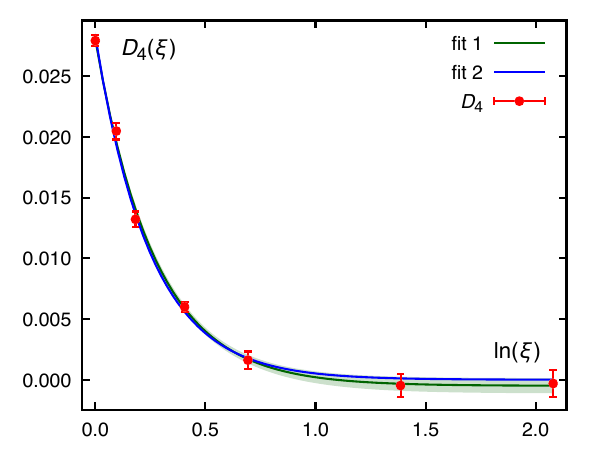}\hfill
\includegraphics[width=0.49\textwidth]{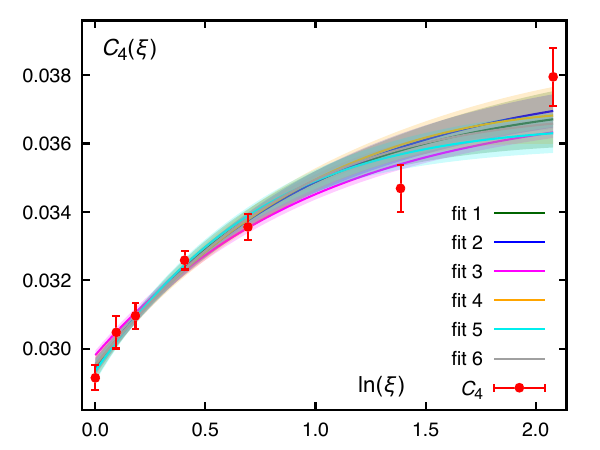}
\caption{The low-energy constants $D_4$ (left) and $C_4$ (right) determined from the aHISQ pion taste splittings as functions of $\ln(\xi)$. The bands represent the fits described in the text.
\label{fig_D4C4_ahisq}}
\end{figure}

Our next step is to determine what functional forms may describe the anisotropy dependence of the low-energy constants $C_4(\xi)$ and $D_4(\xi)$. 
At the first stage, we fit them individually and do not enforce the constraint $D_4(\xi=1)=C_4(\xi=1)$. That constraint is automatically incorporated in the final global fit.
It appears that the $D_4$ data strongly favors a power law. A fit to the form ($D_4$ fit 1):
\begin{equation}
D_4(\xi)=d_1+\frac{d_2}{\xi^{d_3}}
\end{equation}
produces $d_1=-0.00050(62)$, $d_2=0.02855(72)$ and $d_3=3.70(23)$ with $\chi^2/\mbox{dof}=0.9$. 
As the closest integer power is $d_3=4$ and $d_1$ is consistent with 0, as expected, a fit to a simplified form
($D_4$ fit 2):
\begin{equation}
D_4(\xi)=\frac{d_2}{\xi^{4}}
\end{equation}
gives $d_2=0.02830(40)$ with $\chi^2/\mbox{dof}=0.9$. 
The two fits described above are shown in Fig.~\ref{fig_D4C4_ahisq}, left. 
We note that fits to the form
\begin{equation}
D_4(\xi)=d_1+d_2\,\frac{\ln(\xi)}{\xi^{d_3}}
\end{equation}
are disfavored by the data, with $\chi^2/\mbox{dof}$ of order ten.
 
Next, we examine the dependence $C_4(\xi)$ where we find the data to be less constraining. A fit to the form
($C_4$ fit 1):
\begin{equation}
C_4(\xi)=c_1\,\left(1-\frac{c_2}{\xi^{c_3}}\right)
\end{equation}
produces $c_1=0.0375(12)$, $c_2=0.215(24)$ and $c_3=1.14(31)$ with $\chi^2/\mbox{dof}=1.6$. 
The power is consistent with 1, so we fix $c_3=1$, then the fit
($C_4$ fit 2)
produces $c_1=0.03802(51)$, $c_2=0.225(15)$ with $\chi^2/\mbox{dof}=1.3$. 
Fixing also $c_2=1/5$
($C_4$ fit 3)
gives $c_1=0.03726(18)$ with $\chi^2/\mbox{dof}=1.5$.
In general, we also expect some logarithmic dependence on $\xi$ (if perturbation theory is applicable, one expects the coupling also to appear in the expansion). We therefore also try
($C_4$ fit 4):
\begin{equation}
C_4(\xi)=c_1+c_2\,\frac{\ln(\xi)}{\xi^{c_3}}
\end{equation}
that gives $c_1=0.02951(29)$, $c_2=0.0081(12)$ and $c_3=0.403(92)$ with $\chi^2/\mbox{dof}=1.9$. 
We interpret the appearance of non-integer powers of $\xi$ as masking some more complicated dependence, still resolvable by the data, and try
($C_4$ fit 5):
\begin{equation}
C_4(\xi)=c_1\left(1-(c_2+c_3\ln(\xi))\,\frac{1}{\xi^{2}}\right)
\end{equation}
that produces $c_1=0.03658(68)$, $c_2=0.198(16)$ and $c_3=0.140(78)$ with $\chi^2/\mbox{dof}=1.9$. 

Fixing $c_2=c_3=1/5$ ($C_4$ fit 6) gives $c_1=0.03694(18)$ with $\chi^2/\mbox{dof}=1.4$.
The six fit forms described above together with the data are shown in Fig.~\ref{fig_D4C4_ahisq} (right). We give preference to the fits that contain even powers, and possibly logarithms, of the anisotropy $\xi$.

\begin{SCfigure}
\includegraphics[width=0.5\textwidth]{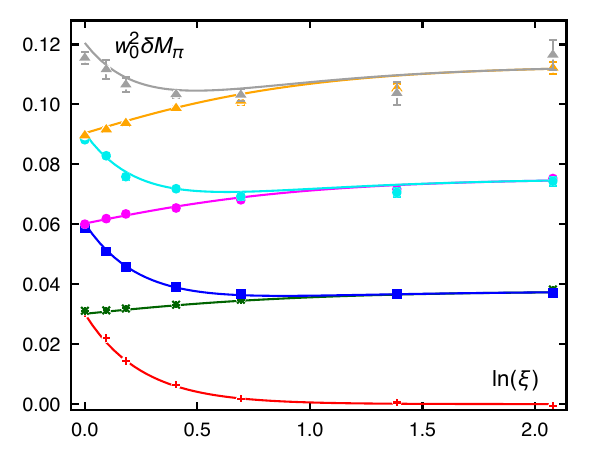}
\caption{Splittings of the aHISQ pion taste multiplets from the Goldstone pion, same as in Fig.~\ref{fig_pion_split} (right), but in units of $1/w_0^2$, as function of $\ln(\xi)$. The lines represent a joint fit of all data points to the functional form in Eq.~(\ref{eq_dM_fit_form}) with a single fit parameter $c$ (overall amplitude).
\label{fig_aHISQ_split_fit}}
\end{SCfigure}

Now we combine the fit forms for $C_4$ and $D_4$ into a single form that we use to fit the aHISQ splittings directly:
\begin{equation}
\label{eq_dM_fit_form}
\delta M_\pi(\xi)=c\left(
l_\sigma\cdot\frac{5}{4}\left(
1-\frac{1}{5}\left(1+\ln(\xi)
\right)
\frac{1}{\xi^2}
\right)
+l_\tau\cdot\frac{1}{\xi^4}
\right),
\end{equation}
where $l_\sigma$ and $l_\tau$ as before are, respectively, the number of unbalanced spatial and temporal links in the interpolating operators.
The $5/4$ coefficient in the first term ensures that in the isotropic limit $\xi=1$ the expression reduces to $c(l_\sigma+l_\tau)$. The remaining fit parameter $c$ is determined by simultaneously fitting all aHISQ splittings to Eq.~(\ref{eq_dM_fit_form}) which gives $c=0.030129(31)$ and $\chi^2/\mbox{dof}=3.2$. 
While the $\chi^2/\mbox{dof}$ is somewhat high, the functional form (\ref{eq_dM_fit_form}) captures the main features of the aHISQ splittings remarkably well, as shown in Fig.~\ref{fig_aHISQ_split_fit}. One of the factors contributing to the high $\chi^2/\mbox{dof}$ is the fact that the model inherits some inaccuracy present already in the isotropic case: Eqs.~(\ref{eq_anichipt_split_noDs_first_hisq})--(\ref{eq_anichipt_split_noDs_last_hisq}) that reduce to the known forms of Ref.~\cite{Lee:1999zxa} predict that the isotropic splittings for the axial, tensor, vector and singlet taste multiplets form ratios 1:2:3:4, while the data show a slightly different pattern 1:1.94:2.92:3.82 which is due to the small but non-negligible dependence on the $C_1$, $C_3$ and $C_6$ low-energy constants.

The $\chi^2/\mbox{dof}$ can be brought close to 1 by introducing different amplitudes $c$ for different splittings and/or introducing fit parameters in the parentheses in Eq.~(\ref{eq_dM_fit_form}). We, however, prefer the simplest fit with minimal number of free parameters.

Curiously, one can arrive at an empirical model that captures main features of the aHISQ taste splittings from considerations very different from the staggered chiral perturbation theory inspired approach described above. (In fact, we found that model first, and that allowed us to introduce proper modifications to the chiral perturbation theory-inspired model.) As the origin and the quantitative dependence of the taste splittings on the lattice spacing, and in our case, anisotropy, are not fully understood, we believe, it is useful for future studies to document those considerations here.

Let us first consider introducing the anisotropy $\xi>1$ as a symmetry-breaking perturbation. We can estimate how it affects the fluctuations of spatial and temporal links. Consider plaquettes normalized to the value at $\xi=1$: $Q_{\mu\nu}=P_{\mu\nu}(\xi)/P_{\mu\nu}(\xi=1)$. Similar to tadpole factors, we can define $u_\sigma(\xi)=\sqrt[4]{Q_{\sigma\sigma}}$ and $u_\tau(\xi)=\sqrt[4]{Q_{\sigma\tau}^2/Q_{\sigma\sigma}}$ as proxies for average spatial and temporal link fluctuations, respectively. For $\xi>1$ $u_\tau(\xi)$ is growing, as temporal links fluctuate less and get closer to $1$, while $u_\sigma(\xi)$ is decreasing, as the spatial links fluctuate more and get farther from $1$. 
Using the data from Table~\ref{tab_pg_plaq} one finds that (at least, for small anisotropies, $1 < \xi\lesssim 2$) $u_\tau(\xi)-1\approx 3.7(1-u_\sigma(\xi))$, \textit{i.e.}, per one unit of increase in fluctuations in the spatial links, there are 3.7 units of decrease in fluctuations in the temporal links. 
While the prefactor is not exactly 3 (the number of spatial dimensions), it is plausible to assume a picture similar to how a symmetry-breaking perturbation often affects the spectrum, by redistributing the energy according to multiplicities of the levels with the net change being 0.

This picture is generally familiar, from Grotrian diagrams of fine structure and hyperfine structure in atomic physics, or from the eightfold way~\cite{Gell-Mann:1964ewy}: the larger strange quark mass (or, more precisely, $(m_s-m_l)/(2m_l+m_s) \simeq 0.90$) acts as a symmetry breaking perturbation from the SU(3) flavor symmetric point and, for instance, the baryon decuplet masses split according to their multiplicities in such a way that the weighted average mass stays invariant (up to per mill level amount of about 2~MeV due to the isospin breaking effects). The $\Delta$ state masses are lowered and the $\Xi^*$ and $\Omega$ are raised, while the $\Sigma^*$ states stay at the symmetric point. 
The analog to the symmetry breaking by the strange quark mass for the anisotropic case is that one temporal link getting smoother is exactly balanced by three spatial links getting rougher.

Next, the aHISQ splittings in Fig.~\ref{fig_pion_split} exhibit approximately equidistant pattern and they are ordered by the number of unbalanced links in the interpolating operator. Let the first splitting between the Goldstone and the axial vector taste be denoted $\delta E$. Then
\begin{equation}
\label{eq_dM_iso}
\delta M_\pi=
(l_\sigma+l_\tau)\,\delta E.
\end{equation}
We stress that this formula does not depend on the structure of Eqs.~(\ref{eq_anichipt_split_noDs_first_hisq})--(\ref{eq_anichipt_split_noDs_last_hisq}) that we discussed previously and follows from the aHISQ data alone (\textit{i.e.}, equidistant multiplets and ordering by the number of unbalanced links).
Let us assume that $\delta M_s$ is independent of anisotropy and equal to the value at $\xi=1$, then 
$\delta E=\delta M_s/4$.
(The deviation of $\delta M_s$ from constant is about 17\%, Fig.~\ref{fig_pion_split}). How can anisotropy be introduced into Eq.~(\ref{eq_dM_iso}) from the perspective of a symmetry-breaking perturbation? We can view the taste-breaking effects as the amount of energy associated with the individual fluctuations of the unbalanced gauge links in the interpolating operators. After all, the number of unbalanced links is the number of additional links that an anti-quark needs to hop relative to the quark in a pion taste heavier than the Goldstone taste. This picture of independent link fluctuations immediately explains why the isotropic HISQ splittings are ordered the way they are: axial vector taste (1 unbalanced link), tensor (2), vector (3) and singlet (4 unbalanced links). (And this argument largely applies also to the naive staggered case, since, as Fig.~\ref{fig_chipt_lec} shows, the naive splittings are dominated by the $C_4$ low-energy constant.)

If we assume that the symmetry-breaking perturbation for the gauge links can be described by some positive increasing universal function $f(\xi)$ with the limiting behavior $f(\xi\to1)\to0$ and $f(\xi\to\infty)\to1$, and that the ``units of fluctuations'' distribute themselves between the temporal and spatial links as 3:1, it is natural to modify Eq.~(\ref{eq_dM_iso}) in the following way:
\begin{equation}
\label{eq_dM_aniso}
\delta M_\pi(\xi)=
\left(l_\sigma\left(
1+\frac{1}{3}f(\xi)
\right)+l_\tau\left(
1-f(\xi)
\right)\right)\,\delta E.
\end{equation}
The way $f(\xi)$ is defined, the $1-f(\xi)$ in the temporal part shows that the temporal links are becoming smoother with $\xi$ growing and contribute less fluctuations, while $1+f(\xi)/3$ in the spatial part reflects the opposite: the spatial links are getting rougher and contribute more fluctuations, in proportion 1:3 relative to the temporal links. In this model, the taste splittings based on Eq.~(\ref{eq_dM_aniso}) are
\begin{eqnarray}
\delta M_{45}(\xi) &=&
\frac{3}{3}\,
\left(1-f(\xi)\right)\delta E, \label{eq_ani_first}
\\
\delta M_{i5}(\xi) &=&
\frac{1}{3}\,
\left(3+f(\xi)\right)\delta E,
\\
\delta M_{ij}(\xi) &=&
\frac{2}{3}\,
\left(3-f(\xi)\right)\delta E,
\\
\delta M_{i4}(\xi) &=&
\frac{2}{3}\,
\left(3+f(\xi)\right)\delta E,
\\
\delta M_i(\xi) &=&
\frac{1}{3}\,
\left(9-f(\xi)\right)\delta E,
\\
\delta M_4(\xi) &=&
\frac{3}{3}\,
\left(3+f(\xi)\right)\delta E,
\\
\delta M_s(\xi) &=&
\frac{3}{3}\,\,
4\delta E.
\label{eq_ani_last}
\end{eqnarray}
This model captures the crucial feature of the aHISQ taste splittings observed in the data: the masses of the $\xi_i\xi_5$, $\xi_i\xi_4$ and $\xi_4$ tastes increase with anisotropy, the splittings differ only by a multiplicative factor, in proportion 1:2:3, and $\delta M_{45}\to 0$ as $\xi\to \infty$. If one forms ratios that cancel $\delta E$ and isolate $f(\xi)$:
\begin{equation}
f(\xi)=
3\,\frac{\delta M_{i5}(\xi)-\delta M_{45}(\xi)}{3\delta M_{i5}(\xi)+\delta M_{45}(\xi)}
=
3\,\frac{\delta M_{i4}(\xi)-\delta M_{ij}(\xi)}{\delta M_{i4}(\xi)+\delta M_{ij}(\xi)}
=
9\,\frac{\delta M_{4}(\xi)-\delta M_{i}(\xi)}{\delta M_{4}(\xi)+3\delta M_{i}(\xi)},
\end{equation}
one indeed finds data collapse confirming that $f(\xi)$ is a universal function. This feature is independent from the assumption of $\delta E$ being constant. Therefore, to accommodate the observed dependence of $\delta E$ on anisotropy, one can form linear combinations that cancel $f(\xi)$ and allow one to study the isolated dependence $\delta E(\xi)$.

At this stage, however, it is more convenient to connect the qualitative model to the staggered chiral perturbation theory inspired model, as the latter separates the anisotropy dependence of the taste splitting more cleanly. Matching Eqs.~(\ref{eq_ani_first})--(\ref{eq_ani_last}) with Eqs.~(\ref{eq_anichipt_split_noDs_first_hisq})--(\ref{eq_anichipt_split_noDs_last_hisq}) one finds that
\begin{equation}
\delta E(\xi)=\frac{1}{4}(3C_4(\xi)+D_4(\xi))
\end{equation}
and
\begin{equation}
f(\xi)=3\,\frac{C_4(\xi)-D_4(\xi)}{3C_4(\xi)+D_4(\xi)},
\end{equation}
so that $f(\xi)$ is indeed a universal increasing function with proper limits.

Another interesting observation for aHISQ is that as $\delta M_s(\xi=8)\simeq \delta M_s(\xi=1)$, the five approximately equidistant taste multiplets (counting also the Goldstone $\gamma_5\otimes\xi_5$ taste for which by definition $\delta M_5=0$) at $\xi=1$ rearrange themselves at $\xi=8$ into four approximately equidistant multiplets with the unit of splitting being $4/3\,\delta E$, Fig.~\ref{fig_pion_split} (right), as the qualitative model predicts: Eqs.~(\ref{eq_ani_first})--(\ref{eq_ani_last}) with $f(\xi\to\infty)\to1$ give the pattern $\{0,1,1,2,2,3,3\}\times 4/3\,\delta E$. (This is also not at odds with the limiting behavior of Eq.~(\ref{eq_dM_fit_form}), if one observes that at the largest anisotropy $\xi=8$ $\delta E$ has a smaller value: taking $\delta E\to 15/16\,\delta E$ would produce a unit of splitting $5/4\,\delta E$, as dictated by Eq.~(\ref{eq_dM_fit_form}).)
If $\xi=8$ can be considered large enough to represent the $\xi\to\infty$ limit, and whether this behavior of $\delta E$ (\textit{i.e.}, $\delta M_s$) is similar at smaller spatial lattice spacing $a_\sigma$ remains to be uncovered.

\section{Conclusion}
\label{sec_concl}
In this paper we have studied tuning of the parameters of anisotropic gauge and staggered fermion actions in preparation for fully dynamical simulations with the anisotropic highly improved staggered quark (aHISQ) action, introduced in Eq.~(\ref{eq_ahisq}). To minimize the computational cost (which is substantial, as one needs to map out a four-dimensional parameter space), we performed a quenched study. In that case, the gauge, $(\beta,\xi_0)$, and fermion, $(a_\sigma m,\xi_0^f)$ action parameters are tuned independently.

For tuning the bare gauge anisotropy $\xi_0$ we use gradient flow, as described in Ref.~\cite{Borsanyi:2018srz}, which is a slight modification of the earlier procedure of Ref.~\cite{Borsanyi:2012zr}. We studied how different choices of the flow and observable affect tuning and concluded that Wilson flow with clover observable are optimal for our purposes. This gradient flow setup is identical to the one in Ref.~\cite{Borsanyi:2018srz} and our results independently reproduce the dependencies $w_0/a_\sigma(\beta)$ and $\xi_0(\beta)$ (that we call lines of constant renormalized anisotropy, LCRA) for the renormalized anisotropy $\xi=2$ used in that work in the range of couplings that we studied.

With the gradient flow, we tuned a set of ensembles with the spatial lattice spacing $a_\sigma\simeq0.1665$~fm that span a range of renormalized anisotropies $\xi=1,\dots,8$. On those ensembles we studied simultaneous tuning of the bare strange quark mass and bare fermion anisotropy. The renormalized fermion anisotropy is defined through the meson dispersion relation, although we also tested using the definition from the ratio of ground state energies in the temporal and spatial correlation functions. We have tuned the naive staggered action in the range of anisotropies $\xi=1,\dots,2$ and the aHISQ action in the range $\xi=1,\dots,8$. While the tuning of the naive action requires a two-dimensional scan in the $(a_\sigma m_s,\xi_0^f)$ parameter space, the aHISQ one can reduce to two one-dimensional scans.

After tuning the strange quark mass and bare fermion anisotropy for the naive and aHISQ action we measured the pion taste masses and splittings with the light quark mass set to one fifth of the strange (this corresponds to the mass of the lowest, Goldstone pion taste of about 300~MeV). As discussed in Sec.~\ref{sec_pion_split}, we observe qualitatively different behavior of the splittings with anisotropy for the two actions. Attempting to explain and quantify the dependence of the splittings (mainly for aHISQ where: a) we have a larger available anisotropy range, and b) due to suppression of some of the taste-breaking operators the pattern of splittings is simpler than for naive) on anisotropy we developed 
an empirical model for taste splittings. We arrived at the model from two very different starting points: staggered chiral perturbation theory inspired ansatz and the independent link fluctuations picture with anisotropy acting as a symmetry-breaking perturbation. 
They are discussed in Sec.~\ref{sec_schipt}.
While these models describe the aHISQ taste mass splittings remarkably well, Fig.~\ref{fig_aHISQ_split_fit}, further study with multiple anisotropies and spatial lattice spacings is required to unambiguously determine the dependence of the taste splittings on anisotropy and lattice spacing, in general.

In preparation for dynamical aHISQ simulations, we spelled out an algorithm for the aHISQ force calculation in Sec.~\ref{sec_dyn_ani}. To validate our code we reproduced some pure gauge~\cite{CP-PACS:2001lwl}, quenched naive staggered spectrum~\cite{Nomura:2004qu}, and naive staggered dynamical~\cite{Levkova:2006gn} results from the literature, discussed in Appendix~\ref{sec_app_tests}. Additional tests specifically for the aHISQ fermion force are discussed in Appendix~\ref{sec_app_hisq_force}.

The two main directions for the future research are: a) a comprehensive study of the pion taste spectrum that requires more extensive coverage of the $(a_\sigma,\xi)$ parameters space (and a lot can still be learned from computationally cheaper quenched simulations), and b) tuning of the lines of constant renormalized anisotropy in the four-dimensional dynamical aHISQ parameter space $(\beta,\xi_0,a_\sigma m_s, \xi_0^f)$ (where our present finding, $\xi_0^f\simeq\xi$ should lead to significant computational savings).

\acknowledgments

We thank Thomas Chuna for the help at the very early stages of the project, Claude Bernard for discussions on staggered chiral perturbation theory, Maarten Golterman, Jan Smit and Aaron Meyer for discussions on the properties and group theory of staggered fermions, Szabolcs Borsanyi for communicating the data used in Ref.~\cite{Borsanyi:2018srz}, Hwancheol Jeong and William Jay for consultations on \texttt{lsqfit} and Claude Bernard, Szabolcs Borsanyi and Leon Hostetler for careful reading and comments on the manuscript.

A.B and Y.T.’s research is funded by the U.S. National Science Foundation under 
the award No. PHY-2309946.

The lattice QCD calculations have been performed using the publicly available \href{https://web.physics.utah.edu/~detar/milc/milcv7.html}{MILC code}. Computational resources used in this work were in part provided by the Institute for Cyber-Enabled Research at Michigan State University, USQCD Collaboration, funded by the Office of Science of the U.S. Department of Energy, and the National Energy Research Scientific Computing Center (NERSC), a U.S. Department of Energy Office of Science User Facility located at Lawrence Berkeley National Laboratory, operated under Contract
No. DE-AC02-05CH11231.

\appendix

\section{Pure gauge ensembles}
\label{sec_app_ens}
The parameters of the production pure gauge ensembles used for the spectrum studies are documented in Table~\ref{tab_pg_ens}. The values of the spatial and temporal plaquettes on those ensembles in the normalization $P(\beta\to\infty)=1$ are documented in Table~\ref{tab_pg_plaq}.

\setlength{\tabcolsep}{9pt}
\begin{table}
\begin{center}
\caption{Pure gauge ensembles used in this study, tuned to have the same spatial lattice spacing $a_\sigma\approx 0.1665$~fm in the WC scheme. The first column lists the bare gauge coupling $\beta=10/g_0^2$, second the bare fermion anisotropy $\xi_0$, third the target renormalized anisotropy $\xi$, fourth the lattice volume, fifth the number of configurations used in the flow measurements, sixth the ratio of the spatial and temporal scales which is equal to 1 if the ensemble is properly tuned, seventh the gradient flow scale in units of the spatial lattice spacing and eighth the lattice spacing in fm.
\label{tab_pg_ens}
}
\begin{tabular}{llllrlll}
\toprule
\toprule
$\beta$ & $\xi_0$ & $\xi$ &  $N_\sigma^3\times N_\tau$ & $N_{flow}$ &
$w_{0,\sigma}/w_{0,\tau}$ &  $w_{0,\sigma}/a_\sigma$ & $a_\sigma$,~fm  \\
\midrule
6.81823 & 1 & 1 & $16^3\times32$ & 500 & 1 &
1.04147(49) & 0.166639(78) \\
6.84652 & 1.07865 & 1.1 & $16^3\times32$ &300 & 0.99944(26) &
1.04393(87) & 0.16625(14) \\
6.87348 & 1.15792 & 1.2 & $16^3\times32$ & 600 & 0.99946(30) &
1.04411(41) & 0.166218(65) \\
6.94635 & 1.39939 & 1.5 & $16^3\times48$ & 500 & 1.00080(43) &
1.04090(56) & 0.166731(90) \\
7.04115 & 1.81411 & 2 & $16^3\times64$ & 440 & 0.99970(34) &
1.04158(79) & 0.16662(13) \\
7.19156 & 3.48992 & 4 & $16^3\times128$ & 1640 &  1.00043(17) &
1.04165(32) & 0.166611(51) \\
7.26025 & 6.89327 & 8 & $20^3\times320$ & 320 & 0.99897(25) &
1.03859(78) &  0.16710(13) \\
\bottomrule
\bottomrule
\end{tabular}
\end{center}
\end{table}

\setlength{\tabcolsep}{9pt}
\begin{table}
\begin{center}
\caption{
Spatial, $P_{\sigma\sigma}$ (column 5), and temporal, $P_{\sigma\tau}$ (column 6), plaquettes computed on the production pure gauge ensembles. The meaning of the columns 1--3 is the same as in Table~\ref{tab_pg_ens}, column 4 lists the number of plaquette measurements.
\label{tab_pg_plaq}
}
\begin{tabular}{lllrll}
\toprule
\toprule
$\beta$ & $\xi_0$ & $\xi$ & $N_{meas}$ & $P_{\sigma\sigma}$ &
$P_{\sigma\tau}$ \\
\midrule
6.81823 & 1 & 1 & 80,000 & 0.5815390(34) & 0.5815417(32) \\
6.84652 & 1.07865 & 1.1 & 80,000 & 0.5663272(36) & 0.6023595(29) \\
6.87348 & 1.15792 & 1.2 & 80,000 & 0.5528877(50) & 0.6207760(34) \\
6.94635 & 1.39939 & 1.5 & 80,000 & 0.5214717(30) & 0.6662955(16) \\
7.04115 & 1.81411 & 2 & 40,000 & 0.4892744(43) & 0.7222637(18) \\
7.19156 & 3.48992 & 4 & 30,000 & 0.4456197(53) & 0.83347145(60) \\
7.26025 & 6.89327 & 8 & 15,000 & 0.4282994(20) & 0.90823420(26) \\
\bottomrule
\bottomrule
\end{tabular}
\end{center}
\end{table}

\section{Spectrum fits}
\label{sec_app_spec_fits}
In this appendix we document the results of the fits used to tune the bare fermion anisotropy $\xi_0^f$ for the aHISQ action, as well as the pion taste masses computed on the pure gauge ensembles listed in Table~\ref{tab_pg_ens}. As discussed in the main text, we used full two-dimensional grid approach to tune the renormalized fermion anisotropy $\xi^f=1.2$ and $1.5$. The results for the energy levels are documented in Tables~\ref{tab_tune_ahisq_xi12} and \ref{tab_tune_ahisq_xi15}.

For the $\xi^f=1.1$, $2$, $4$ and $8$ we used a simplified tuning method: one-dimensional search in bare fermion anisotropy at fixed quark mass, then one-dimensional search in quark mass at fixed anisotropy. The results are collected in Tables~\ref{tab_tune_ahisq_xi11}--\ref{tab_tune_ahisq_xi8}. Those tables contain two parts, with the meaning of the columns being different to accommodate finite momentum fits in the first and the mass dependence in the second.

The predicted values of $(a_\sigma m,\xi_0^f)$ for the tuned ensembles are summarized in Table~\ref{tab_splittings_run_params} in columns five and six under the ``predicted'' header. The actual run parameters, listed in columns seven and eight are discussed in Appendix~\ref{sec_app_quality}.

The masses of different pion tastes for the aHISQ action are shown in Table~\ref{tab_taste_mass_ahisq} and the quadratic mass splittings from the Goldstone taste $\xi_5$ in Table~\ref{tab_taste_split_ahisq}. Each line of that table corresponds to the same line in Table~\ref{tab_pg_ens}. The number of configurations used for these measurements for $\xi=1$ is 4,400, for $\xi=1.1$ 4,200, for $\xi=1.2$ 4,000, for $\xi=1.5$ 4,650, for $\xi=2$ 2,000, for $\xi=4$ 1,600 and for $\xi=8$ 800.

The masses and the quadratic splittings for the naive staggered action are shown in Tables~\ref{tab_taste_mass_naive} and \ref{tab_taste_split_naive}, respectively. The number of configurations used for $\xi=1$ is 4,400, for $\xi=1.2$ 4,850, for $\xi=1.5$ 4,650 and for $\xi=2$ 2,200.

\setlength{\tabcolsep}{9pt}
\begin{table}
\begin{center}
\caption{
Tuning of the strange quark mass $a_\sigma m_s$ for the aHISQ action on the isotropic pure gauge ensemble $(\beta=6.81823,\xi_0=1)$ listed in Table~\ref{tab_pg_ens}. The first row lists the strange quark mass $a_\sigma m_s$ and second the energy level at zero momentum $a_\tau E(0)$, \textit{i.e.} the mass of $\eta_{s\bar s}$, measured in units of temporal lattice spacing.
\label{tab_tune_ahisq_xi1}
}
\begin{tabular}{l | lllll}
\toprule
\toprule
$a_\sigma m_s $ & 0.03 & 0.05 & 0.07 & 0.09 & 0.11 \\
$a_\tau E(0)$ & 0.35186(47) & 0.45084(43) & 0.53190(41) &
0.60299(41) & 0.66761(41) \\
\bottomrule
\bottomrule
\end{tabular}
\end{center}
\end{table}

\setlength{\tabcolsep}{9pt}
\begin{table}
\begin{center}
\caption{
Tuning of the bare fermion anisotropy $\xi_0^f$ and the strange quark mass $a_\sigma m_s$ for the aHISQ action on the pure gauge ensemble $(\beta=6.87348,\xi_0=1.15792)$ with the renormalized anisotropy $\xi=1.2$, listed in Table~\ref{tab_pg_ens}. The first column lists the bare fermion anisotropy $\xi_0^f$, second the strange quark mass $a_\sigma m_s$, third, fourth and fifth the energy $a_\tau E(\vec{n}^2)$ extracted from the fits of correlation functions at the momentum $\vec{n}$ in lattice units, sixth the renormalized fermion anisotropy as defined by the dispersion relation, Eq.~(\ref{eq_disprel}). The number of configurations used for these measurements $N_{conf}=950$.
\label{tab_tune_ahisq_xi12}
}
\begin{tabular}{llllll}
\toprule
\toprule
$\xi_0^f$ & $a_\sigma m_s$ & $a_\tau E(0)$ &
$a_\tau E(1)$ &  $a_\tau E(2)$ & $\xi^f$  \\
\midrule
1 & 0.05 & 0.41582(31) & 0.55032(67) & 0.65957(201) & 1.08821(265) \\
& 0.07 & 0.49201(32) & 0.61089(58) & 0.71131(146) & 1.08332(227) \\
& 0.09 & 0.55949(32) &  0.66687(56) & 0.76027(103) & 1.08046(228) \\
\midrule
1.2 & 0.05 & 0.37596(32) & 0.49735(71) & 0.59564(143) & 1.20437(317) \\
& 0.07 & 0.44364(33) & 0.55034(60) & 0.64046(100) & 1.20390(275) \\
& 0.09 & 0.50315(32) & 0.59925(54) & 0.68256(74) & 1.20487(247) \\
\midrule
1.4 & 0.05 & 0.34569(34) & 0.45661(70) & 0.54617(117) & 1.31483(348) \\
& 0.07 & 0.40704(34) & 0.50411(58) & 0.58548(75) & 1.31990(290) \\
& 0.09 & 0.46063(32) & 0.54773(52) & 0.62301(66) & 1.32426(278) \\
\bottomrule
\bottomrule
\end{tabular}
\end{center}
\end{table}

\setlength{\tabcolsep}{9pt}
\begin{table}
\begin{center}
\caption{
Tuning of the bare fermion anisotropy $\xi_0^f$ and the strange quark mass $a_\sigma m_s$ for the aHISQ action on the pure gauge ensemble $(\beta=6.94635,\xi_0=1.39939)$ with the renormalized anisotropy $\xi=1.5$, listed in Table~\ref{tab_pg_ens}. The meaning of the columns is the same as in Table~\ref{tab_tune_ahisq_xi12}. The number of configurations used for these measurements $N_{conf}=500$.
\label{tab_tune_ahisq_xi15}
}
\begin{tabular}{llllll}
\toprule
\toprule
$\xi_0^f$ & $a_\sigma m_s$ & $a_\tau E(0)$ &
$a_\tau E(1)$ &  $a_\tau E(2)$ & $\xi^f$  \\
\midrule
1.05 & 0.03 & 0.28705(35) & 0.42730(115) &
0.53087(160) & 1.24236(378) \\
& 0.05 & 0.36959(36) & 0.48803(79) & 0.58168(125) & 1.23451(350) \\
& 0.07 & 0.43853(35) & 0.54329(64) & 0.62976(115) & 1.22667(347) \\
& 0.09 & 0.50001(34) & 0.59536(68) & 0.67574(104) & 1.21911(320) \\
\midrule
1.25 & 0.03 & 0.26121(32) & 0.38896(88) & 0.48319(155) & 1.36429(406) \\
& 0.05 & 0.33535(33) & 0.44285(64) & 0.52824(124) & 1.35907(382) \\
& 0.07 & 0.39674(32) & 0.49158(54) & 0.57035(107) & 1.35406(369) \\
& 0.09 & 0.45113(32) & 0.53726(54) & 0.61090(122) & 1.34690(367) \\
\midrule
1.45 & 0.03 & 0.24132(29) & 0.35904(70) & 0.44580(153) & 1.47876(423) \\
& 0.05 & 0.30911(30) & 0.40808(56) & 0.48698(160) & 1.47447(425) \\
& 0.07 & 0.36480(32) & 0.45205(50) & 0.52485(131) & 1.47124(394) \\
& 0.09 & 0.41398(30) & 0.49251(43) & 0.55990(107) & 1.47232(360) \\
\midrule
1.65 & 0.03 & 0.22524(28) & 0.33479(59) & 0.41538(144) & 1.58728(453) \\
& 0.05 & 0.28812(28) & 0.37963(49) & 0.45361(161) & 1.58783(435) \\
& 0.07 & 0.33949(29) & 0.41970(43) & 0.48710(121) & 1.59093(400) \\
& 0.09 & 0.38448(29) & 0.45687(40) & 0.51912(82) &  1.59167(368) \\
\bottomrule
\bottomrule
\end{tabular}
\end{center}
\end{table}

\setlength{\tabcolsep}{9pt}
\begin{table}
\begin{center}
\caption{
Simplified tuning of the bare fermion anisotropy $\xi_0^f$ at fixed strange quark mass (upper part of the table) and the strange quark mass $a_\sigma m_s$ tuning at fixed bare fermion anisotropy (lower part of the table) on the pure gauge ensemble $(\beta=6.84652,\xi_0=1.07865)$ with the renormalized anisotropy $\xi=1.1$, listed in Table~\ref{tab_pg_ens}. The meaning of the columns in the upper part is the same as in Tables~\ref{tab_tune_ahisq_xi12} and \ref{tab_tune_ahisq_xi15}. The lower part contains measurements only at zero momentum and the meaning of the rows is the same as for the isotropic ensemble in Table~\ref{tab_tune_ahisq_xi1} with the additional first column showing the bare fermion anisotropy $\xi_0^f$ at which the measurements were done. The number of configurations used for these measurements is $N_{conf}=400$.
\label{tab_tune_ahisq_xi11}
}
\begin{tabular}{llllll}
\toprule
\toprule
$\xi_0^f$ & $a_\sigma m_s$ & $a_\tau E(0)$ &
$a_\tau E(1)$ &  $a_\tau E(2)$ & $\xi^f$  \\
\midrule
1 & 0.06 & 0.47296(63) & 0.60457(114) & 0.71208(768) & 1.04284(477) \\
1.05 & 0.06 & 0.46020(63) & 0.58816(112) & 0.69302(690) & 1.07214(488) \\
1.1 & 0.06 & 0.44842(63) & 0.57296(111) & 0.67523(618) & 1.10095(502) \\
1.15 & 0.06 & 0.43748(63) & 0.55882(110) & 0.65857(555) & 1.12932(517) \\
\midrule
\midrule
$\xi_0^f$ & \multicolumn{1}{l|}{$a_\sigma m_s$}  & 0.04 & 0.06 & 0.08 & 0.1 \\
1.08 & \multicolumn{1}{l|}{$a_\tau E(0)$} & 0.37166(66) & 0.45302(63) &  0.52254(60) &
0.58488(57) \\
\bottomrule
\bottomrule
\end{tabular}
\end{center}
\end{table}

\setlength{\tabcolsep}{9pt}
\begin{table}
\begin{center}
\caption{
Simplified tuning of the bare fermion anisotropy $\xi_0^f$ at fixed strange quark mass (upper part of the table) and the strange quark mass $a_\sigma m_s$ tuning at fixed bare fermion anisotropy (lower part of the table) on the pure gauge ensemble $(\beta=7.04115,\xi_0=1.81411)$ with the renormalized anisotropy $\xi=2$, listed in Table~\ref{tab_pg_ens}. The layout of the table is the same as in Table~\ref{tab_tune_ahisq_xi11}. The number of configurations used for the measurements in the upper part is $N_{conf}=400$, and in the lower part $N_{conf}=500$.
\label{tab_tune_ahisq_xi2}
}
\begin{tabular}{lllllll}
\toprule
\toprule
$\xi_0^f$ & $a_\sigma m_s$ & $a_\tau E(0)$ &
$a_\tau E(1)$ &  $a_\tau E(2)$ & $\xi^f$  & - \\
\midrule
1.84 & 0.07 & 0.28287(29) & 0.34986(62) & 0.40691(123) & 1.90339(784) & -\\
1.92 & 0.07 & 0.27622(28) & 0.34184(55) & 0.39717(113) & 1.94816(731) & -\\
2 & 0.07 & 0.27001(28) & 0.33404(52) & 0.38803(105) & 1.99493(711) & -\\
\midrule
\midrule
$\xi_0^f$ & \multicolumn{1}{l|}{$a_\sigma m_s$} & 0.03 & 0.05 & 0.07 & 0.09 & 0.11 \\
1.98  & \multicolumn{1}{l|}{$a_\tau E(0)$} & 0.17986(25) & 0.23016(27) & 0.27165(25) &
0.30815(24) & 0.34145(23) \\
\bottomrule
\bottomrule
\end{tabular}
\end{center}
\end{table}

\setlength{\tabcolsep}{9pt}
\begin{table}
\begin{center}
\caption{
Simplified tuning of the bare fermion anisotropy $\xi_0^f$ at fixed strange quark mass (upper part of the table) and the strange quark mass $a_\sigma m_s$ tuning at fixed bare fermion anisotropy (lower part of the table) on the pure gauge ensemble $(\beta=7.19156,\xi_0=3.48992)$ with the renormalized anisotropy $\xi=4$, listed in Table~\ref{tab_pg_ens}. The layout of the table is the same as in Table~\ref{tab_tune_ahisq_xi11}. The number of configurations used for these measurements is $N_{conf}=500$.
\label{tab_tune_ahisq_xi4}
}
\begin{tabular}{lllllll}
\toprule
\toprule
$\xi_0^f$ & $a_\sigma m_s$ & $a_\tau E(0)$ &
$a_\tau E(1)$ &  $a_\tau E(2)$ & $\xi^f$  & - \\
\midrule
3.76 & 0.05 & 0.11985(12) & 0.15803(50) & 0.18671(107) & 3.83782(2262) & - \\
3.88 & 0.05 & 0.11783(12) & 0.15530(45) & 0.18355(100) & 3.90586(2177)  & - \\
4 & 0.05 & 0.11595(13) & 0.15269(42) & 0.18053(94) & 3.97497(2044) & -\\
\midrule
\midrule
$\xi_0^f$ & \multicolumn{1}{l|}{$a_\sigma m_s$} & 0.03 & 0.05 & 0.07 & 0.09 & 0.11 \\
4 & \multicolumn{1}{l|}{$a_\tau E(0)$} & 0.09060(15) & 0.11595(13) & 0.13682(12) &
0.15520(12) & 0.17190(11) \\
\bottomrule
\bottomrule
\end{tabular}
\end{center}
\end{table}

\setlength{\tabcolsep}{9pt}
\begin{table}
\begin{center}
\caption{
Simplified tuning of the bare fermion anisotropy $\xi_0^f$ at fixed strange quark mass (upper part of the table) and the strange quark mass $a_\sigma m_s$ tuning at fixed bare fermion anisotropy (lower part of the table) on the pure gauge ensemble $(\beta=7.26025,\xi_0=6.89327)$ with the renormalized anisotropy $\xi=8$, listed in Table~\ref{tab_pg_ens}. The layout of the table is the same as in Table~\ref{tab_tune_ahisq_xi11}. The number of configurations used for these measurements is $N_{conf}=300$.
\label{tab_tune_ahisq_xi8}
}
\begin{tabular}{llllll}
\toprule
\toprule
$\xi_0^f$ & $a_\sigma m_s$ & $a_\tau E(0)$ &
$a_\tau E(1)$ &  $a_\tau E(2)$ & $\xi^f$ \\
\midrule
7.6 & 0.07 & 0.070646(99) & 0.08128(14) & 0.09097(33) & 7.78876(4601) \\
7.8 & 0.07 & 0.069630(95) & 0.08013(14) & 0.08969(29) & 7.89090(4691) \\
8 & 0.07 & 0.068640(83) & 0.07900(14) & 0.08841(27) & 8.00298(4682) \\
8.2 & 0.07 & 0.067721(93) & 0.07791(14) & 0.08717(27) & 8.12518(5046) \\
\midrule
\midrule
$\xi_0^f$ &  \multicolumn{1}{l|}{$a_\sigma m_s$} & 0.05 & 0.06 & 0.07 & 0.08 \\
7.87 & \multicolumn{1}{l|}{$a_\tau E(0)$} & 0.058655(94) & 0.064172(92) & 0.069283(87) &
0.074085(83) \\
\bottomrule
\bottomrule
\end{tabular}
\end{center}
\end{table}

\begin{table}
\begin{center}
\caption{Predicted values for the strange quark mass $a_\sigma m_s$ and the bare fermion anisotropy $\xi_0^f$ (columns five and six), and actual values of these parameters (columns seven and eight), used for the splittings measurements. The light quark mass $a_\sigma m_l$ used was equal to one fifth of the actual $a_\sigma m_s$ (column 7). The first column lists the fermion action and the information needed to identify the pure gauge ensembles used, the renormalized anisotropy, gauge coupling and the bare gauge anisotropy is listed in columns two to four, respectively.
\label{tab_splittings_run_params}}
\begin{tabular}{l @{\hspace{0.05\textwidth}} lll @{\hspace{0.05\textwidth}} ll @{\hspace{0.05\textwidth}} ll}
\toprule
\toprule
&&&&\multicolumn{2}{c}{predicted}&\multicolumn{2}{c}{actual}\\
\midrule
action & $\xi$ & $\beta$ & $\xi_0$ & $a_\sigma m_s$ & $\xi_0^f$ & $a_\sigma m_s$ & $\xi_0^f$\\
\midrule
naive & 1   & 6.81823 & 1       & 0.046575(38) & 1    & 0.04275 & 1       \\
& 1.2 & 6.87348 & 1.15792 & 0.040412(44) & 1.0586(28) & 0.0374 & 1.058   \\
& 1.5 & 6.94635 & 1.39939 & 0.040159(38) & 1.2394(38) & 0.0367 & 1.2348  \\
& 2   & 7.04115 & 1.81411 & 0.039816(20) & 1.5577(26) & 0.03625 & 1.54707 \\
\midrule
aHISQ & 1   & 6.81823 & 1       &0.08255(12) & 1      & 0.0762 & 1       \\
& 1.1 & 6.84652 & 1.07865 & 0.08062(18) & 1.0983(88)  & 0.07485 & 1.08    \\
& 1.2 & 6.87348 & 1.15792 & 0.08179(11) & 1.1925(44)  & 0.07605 & 1.195   \\
& 1.5 & 6.94635 & 1.39939 & 0.08150(13) & 1.4949(63)  & 0.0747 & 1.4972  \\
& 2   & 7.04115 & 1.81411 & 0.07917(14) & 2.009(12)   & 0.073 & 1.98    \\
& 4   & 7.19156 & 3.48992 & 0.07805(13) & 4.044(35)   & 0.0723 & 4       \\
& 8   & 7.26025 & 6.89327 & 0.07681(18) & 7.986(85)   & 0.0708 & 7.87    \\
\bottomrule
\bottomrule
\end{tabular}
\end{center}
\end{table}

\setlength{\tabcolsep}{3pt}
\begin{table}
\small
\begin{center}
\caption{
Masses $a_\tau M_\pi$ of the eight pion tastes with taste structure given in the column header, in units of temporal lattice spacing, for the aHISQ action on pure gauge ensembles. Each line corresponds to an ensemble on the same line of Table~\ref{tab_pg_ens}.
\label{tab_taste_mass_ahisq}
}
\begin{tabular}{llllllll}
\toprule
\toprule
$\xi_5$ & $\xi_{4}\xi_{5}$ &$\xi_{i}\xi_{5}$ & $\xi_{i}\xi_{j}$ &
$\xi_{i}\xi_{4}$ & $\xi_{i}$ &$\xi_4$ & $1$ \\
\midrule
0.25480(18) & 0.30465(54) & 0.30600(35) & 0.34512(61) &
0.34681(53) & 0.38242(91) & 0.38409(70) & 0.41399(238) \\
0.23202(14) & 0.26557(40) & 0.27846(29) & 0.30404(64) &
0.31743(76) & 0.34156(86) & 0.35113(81) & 0.37216(335) \\
0.21305(17) & 0.23351(34) & 0.25636(31) & 0.27292(55) &
0.29293(58) & 0.30616(104) & 0.32417(85) & 0.33655(228) \\
0.169883(95) & 0.17755(17) & 0.20608(18) & 0.21184(51) &
0.23602(38) & 0.24153(60) & 0.26335(52) & 0.26695(96) \\
0.12721(14) & 0.12883(18) & 0.15558(26) & 0.15696(46) &
0.17858(51) & 0.17932(75) & 0.19871(76) & 0.19989(113) \\
0.063578(83) & 0.06382(10) & 0.07842(13) & 0.07847(27) &
0.09040(26) & 0.09006(46) & 0.10053(42) & 0.10009(111) \\
0.032570(91) & 0.032456(55) & 0.04018(11) & 0.03996(19) &
0.04638(14) & 0.04624(23) & 0.05182(26) & 0.05243(69) \\
\bottomrule
\bottomrule
\end{tabular}
\end{center}
\end{table}

\setlength{\tabcolsep}{3pt}
\begin{table}
\small
\begin{center}
\caption{
Quadratic mass splittings $a_\tau^2(M_\pi^2-M_5^2)$ of the heavier tastes given in the column headers from the Goldstone taste $\xi_5$, in units of temporal lattice spacing, for the aHISQ action on pure gauge ensembles. Each line corresponds to an ensemble on the same line of Table~\ref{tab_pg_ens}. We attribute the negative value for the $\xi_4\xi_5$ splitting at the largest anisotropy $\xi=8$ (last line) to a fluctuation, as we do not expect any splittings to become negative, and interpret that result as consistent with 0.
\label{tab_taste_split_ahisq}
}
\begin{tabular}{lllllll}
\toprule
\toprule
$\xi_{4}\xi_{5}$ &$\xi_{i}\xi_{5}$ & $\xi_{i}\xi_{j}$ &
$\xi_{i}\xi_{4}$ & $\xi_{i}$ &$\xi_4$ & $1$ \\
\midrule
0.02789(32) & 0.02872(18) & 0.05419(43) & 0.05535(35) & 
0.08132(71) & 0.08260(52) & 0.1065(20)\\
0.01670(21) & 0.02371(14) & 0.03861(39) & 0.04693(48) & 
0.06283(58) & 0.06946(57) & 0.0847(25)\\
0.00914(14) & 0.02033(13) & 0.02909(30) & 0.04042(31) & 
0.04834(63) & 0.05970(53) & 0.0679(15)\\
0.002664(55) & 0.013608(61) & 0.01602(22) & 0.02684(18) & 
0.02948(28) & 0.04050(27) & 0.04240(50)\\
0.000414(43) & 0.008023(65) & 0.00845(14) & 0.01571(18) & 
0.01597(27) & 0.02330(30) & 0.02377(46)\\
0.000031(13) & 0.002107(20) & 0.002116(38) & 0.004130(48) & 
0.004068(82) & 0.006064(87) & 0.00598(22)\\
-0.0000074(58) & 0.000554(10) & 0.000536(15) & 0.001090(14) & 0.001077(22) & 0.001624(27) & 0.001688(73)\\
\bottomrule
\bottomrule
\end{tabular}
\end{center}
\end{table}

\setlength{\tabcolsep}{3pt}
\begin{table}
\small
\begin{center}
\caption{
Masses $a_\tau M_\pi$ of the eight pion tastes with taste structure given in the column header, in units of temporal lattice spacing, for the naive staggered action on pure gauge ensembles. 
The first line with the data corresponds to the ensemble with the renormalized anisotropy $\xi=1$ in Table~\ref{tab_pg_ens}, second to $\xi=1.2$, third to $\xi=1.5$ and fourth to $\xi=4$.
\label{tab_taste_mass_naive}
}
\begin{tabular}{llllllll}
\toprule
\toprule
$\xi_5$ & $\xi_{4}\xi_{5}$ &$\xi_{i}\xi_{5}$ & $\xi_{i}\xi_{j}$ &
$\xi_{i}\xi_{4}$ & $\xi_{i}$ &$\xi_4$ & $1$ \\
\midrule
0.261611(99) & 0.5361(24) & 0.5361(17) & 0.6116(35) & 
0.6183(45) & 0.6604(72) & 0.6607(53) & 0.6936(68)\\
0.214802(96) & 0.4042(35) & 0.4368(16) & 0.4840(38) & 
0.4973(29) & 0.5167(55) & 0.5345(54) & 0.5471(58)\\
0.171252(73) & 0.2870(15) & 0.3375(11) & 0.3576(40) & 
0.3883(32) & 0.3921(23) & 0.4217(29) & 0.4230(49)\\
0.127119(73) & 0.18116(56) & 0.24034(89) & 0.2520(26) & 
0.2817(32) & 0.2852(36) & 0.3104(58) & 0.3097(57)\\
\bottomrule
\bottomrule
\end{tabular}
\end{center}
\end{table}

\setlength{\tabcolsep}{3pt}
\begin{table}
\small
\begin{center}
\caption{
Quadratic mass splittings $a_\tau^2(M_\pi^2-M_5^2)$ of the heavier tastes given in the column headers from the Goldstone taste $\xi_5$, in units of temporal lattice spacing, for the naive staggered action on pure gauge ensembles. The first line with the data corresponds to the ensemble with the renormalized anisotropy $\xi=1$ in Table~\ref{tab_pg_ens}, second to $\xi=1.2$, third to $\xi=1.5$ and fourth to $\xi=4$.
\label{tab_taste_split_naive}
}
\begin{tabular}{lllllll}
\toprule
\toprule
$\xi_{4}\xi_{5}$ &$\xi_{i}\xi_{5}$ & $\xi_{i}\xi_{j}$ &
$\xi_{i}\xi_{4}$ & $\xi_{i}$ &$\xi_4$ & $1$ \\
\midrule
0.2189(26) & 0.2190(19) & 0.3056(43) & 0.3139(55) & 
0.3677(95) & 0.3680(70) & 0.4126(94)\\
0.1173(28) & 0.1446(14) & 0.1881(37) & 0.2012(29) & 
0.2208(56) & 0.2395(58) & 0.2532(64)\\
0.05302(88) & 0.08455(78) & 0.0985(29) & 0.1215(25) & 
0.1244(18) & 0.1485(24) & 0.1496(42)\\
0.01666(20) & 0.04161(43) & 0.0473(13) & 0.0632(18) & 
0.0652(20) & 0.0802(36) & 0.0797(36)\\
\bottomrule
\bottomrule
\end{tabular}
\end{center}
\end{table}

\section{Fermion anisotropy tuning methods}
\label{sec_app_mt_mz}

\begin{SCfigure}
\includegraphics[width=0.5\textwidth]{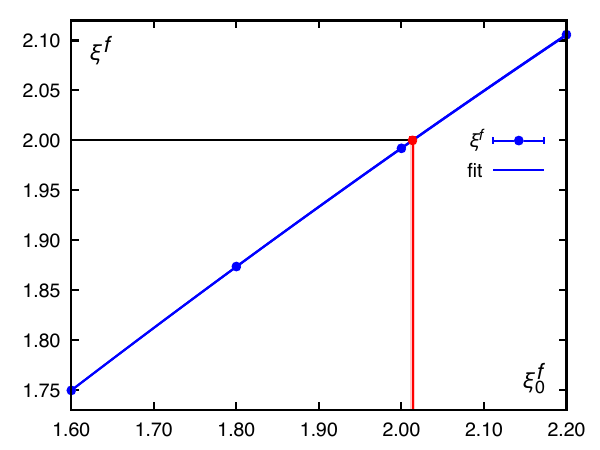}
\caption{Determining the bare fermion anisotropy $\xi_0^f$ that corresponds to the renormalized fermion anisotropy $\xi^f=\xi=2$ with the mass ratio method, where $\xi^f$ is defined as a ratio of masses in lattice units measured in $\tau$ and $z$ direction, Eq.~(\ref{eq_effmassratio}) for the aHISQ action. These measurements were done on a dedicated ensemble with the volume $16^2\times32\times64$ and $(\beta=7.04115,\xi_0=1.81411)$.\label{fig_mass_ratio}}
\end{SCfigure}

We have chosen to use the dispersion relation, Eq.~(\ref{eq_disprel}), as the main method for tuning fermion anisotropy. We have also checked that using the mass ratio method, Eq.~(\ref{eq_effmassratio}), produces essentially same results. On a pure gauge ensemble, specifically generated for the mass ratio tuning with the volume $16^2\times32\times64$ and the renormalized anisotropy $\xi=2$, we have measured the $\eta_{s\bar s}$ correlation functions in the $\tau$ and $z$ direction. The measurements were performed at the strange quark mass $m_s=0.073$ and a set of bare fermion anisotropies $\xi_0^f=1.6$, $1.8$, $2$ and $2.2$. The resulting data: $\xi^f$ defined with Eq.~(\ref{eq_effmassratio}) as function of the bare fermion anisotropy and the quadratic fit are shown in Fig.~\ref{fig_mass_ratio}. The tuned bare fermion anisotropy determined from the fit is $\xi_0^f=2.0134(26)$. This value is comparable with the bare fermion anisotropy predicted with the dispersion relation method: $\xi_0^f=2.009(12)$, and differs from the renormalized anisotropy $\xi=2$ by only $0.7\%$.
We thus expect that the mass ratio tuning method would produce the bare parameters very close to the dispersion relation method. However, we find the latter one to be more convenient for the following reason. Our spatial volumes are typically minimal, to accommodate the gradient flow and to have $m_\pi L\approx 4$. To get stable fits we need the fit direction to be twice larger than the minimal spatial one in physical units. This means that, similar to the example shown in Fig.~\ref{fig_mass_ratio}, for stability of the mass ratio tuning method we need to generate asymmetric spatial volumes $N_z=2N_x=2N_y$ which are, at least, twice more computationally expensive. The dispersion relation tuning method eliminates the need for such asymetric volumes.

\section{Quality of the parameter tuning for the spectrum calculations}
\label{sec_app_quality}
In table~\ref{tab_splittings_run_params} we summarize the parameters predicted from the tuning of the fermion anisotropy and strange quark mass measurements, presented in Tables~\ref{tab_tune_ahisq_xi1}--\ref{tab_tune_ahisq_xi8} and the actual parameters that we used for measuring the pion taste splittings shown in Table~\ref{tab_taste_mass_ahisq}. We remind the reader that we tune the strange quark mass $a_\sigma m_s$ and choose the light quark mass as a fixed ratio, in this case, $1/5$. One can notice that the predicted values differ from the ones we actually used, sometimes as much as about 9\% for the strange quark mass (naive staggered $\xi=2$) and about 1.7\% for the bare fermion anisotropy (aHISQ $\xi=1.1$). Our strange quark mass for both naive and aHISQ is about 7-8\% lower in the actual runs than the predicted values. Here we explain the reasons for that. The overall project whose results we report in this paper did not progress linearly: as we needed many ensembles, gradient flow and spectrum measurements to tune the parameters of the gauge and fermion action, we started multiple simulations, where possible, before full statistics of the tuning runs was collected. Originally, we used a slightly different value of $w_0$ and were aiming at the spatial lattice spacing $a_\sigma=0.16$~fm. As the earlier ensembles turned out to have $a_\sigma\in[0.1662,0.1667]$, we tried to tune the next ensembles closer to $a_\sigma=0.1665$~fm value. However, we already started some spectrum measurements with input $\eta_{s\bar s}$ mass in lattice units based on the expectation that $a_\sigma=0.16$~fm. Thus, our strange quark mass is slightly lower than needed. The pion taste splittings at the lowest order are mass independent, so the mistuning of the mass by a small amount should play no role. (Also, one can reinterpret this situation as that we performed the taste splitting measurements with $a_\sigma m_l\simeq 0.184a_\sigma m_s$ instead of $a_\sigma m_l= 0.2a_\sigma m_s$.)

For the fermion anisotropy we did not expect that $\xi_0^f$ is almost the same as the renormalized anisotropy $\xi$ for the aHISQ action, \textit{e.g.}, Fig.~\ref{fig_xif_Mss_xi12_15_ahisq}. As the expectation was a lower value, similar to the naive staggered action, and it seemed to be confirmed by incomplete statistics, our fermion anisotropy values are also often slightly lower than they should be based on the complete statistics. Overall, the mistuning is quite small, on the level of 1.7\% at most. We would like to make the reader aware of the small discrepancies between the predicted and actual run parameters, but we believe they do not alter the conclusions in this paper.

\section{Comparisons with existing literature}
\label{sec_app_tests}
Our anisotropic setup for calculations in this paper includes the following components: ensemble generation, gradient flow and staggered inverter (naive and aHISQ). To validate our code we reproduced several quantities from the existing literature. In preparation for dynamical aHISQ simulations, we also reproduced some fully dynamical naive staggered simulations.

The pure gauge ensemble generation and gradient flow parts independently reproduce the results of Ref.~\cite{Borsanyi:2018srz} in the relevant range of couplings, \textit{e.g.}, Fig.~\ref{fig_lcra}. To test isolated parts of the code at the early stages of the project we ran various tests. For instance, we generated two pure gauge ensembles with the Wilson gauge action at the parameters of Ref.~\cite{CP-PACS:2001lwl} and compared the plaquette values, Table~\ref{tab_namekawa}.

\setlength{\tabcolsep}{9pt}
\begin{table}
\caption{Comparison of the spatial (first line per $\beta$ value) and temporal (second line) plaquette values with the values from Ref.~\cite{CP-PACS:2001lwl} on two ensembles generated with the Wilson action at the same parameters as in that reference. The first column is the bare gauge coupling $\beta=6/g_0^2$, second the bare gauge anisotropy, third lattice volume, fourth the number of sweeps (one heatbath and four overrelaxation updates of the full lattice), fifth and sixth are the plaquette values and seventh the $Q$ value of the Gaussian difference test.\label{tab_namekawa}}
\begin{tabular}{ l l l l l l l }
\toprule
\toprule
$\beta$ & $\xi_0$ & $N_\sigma^3\times N_\tau$ & $N_{sweep}$ & $P_{\sigma\sigma}$, $P_{\sigma\tau}$ & $P_{\sigma\sigma}$, $P_{\sigma\tau}$ \cite{CP-PACS:2001lwl} & $Q$ \\
\hline
5.8 & 1.67401280 & $16^3\times8$ & 70,000 & 0.462746(41) & 0.462698(75) & 0.57 \\
& & & & 0.688895(18) & 0.688873(33) & 0.56 \\
\hline
6.51881026 & 1.75986308 & $20^3\times40$ & 50,000 & 0.5439713(68) & 0.5439702(40) & 0.89 \\
& & & & 0.7454649(27) & 0.7454657(19) & 0.81 \\
\bottomrule
\bottomrule
\end{tabular}
\end{table}

To compare the naive staggered pion spectrum calculations, we generated a pure gauge ensemble at the same parameters as in Ref.~\cite{Nomura:2004qu} with the Wilson gauge action and the lattice volume of $12^3\times96$. The bare gauge coupling is $\beta=6/g_0^2=5.75$, the bare gauge anisotropy $\xi_0=3.136$, the bare fermion anisotropy $\xi_0^f=2.83$ and the quark mass $a_\sigma m=0.1$. We measured the same pion tastes as in that reference, and the comparison is reported in Table~\ref{tab_nomura}.

\begin{table}
\caption{
Comparison of pion taste masses with Ref.~\cite{Nomura:2004qu}. The parameters of the ensemble are described in the text. The first column lists the taste structure of the measured state, the second column our measurements, the third the results of Ref.~\cite{Nomura:2004qu} and fourth column the $Q$ value of the Gaussian difference test.\label{tab_nomura}}
\begin{tabular}{l l l l}
\toprule
\toprule
taste & $a_\tau M_{\pi}$ & $a_\tau M_{\pi}$  \cite{Nomura:2004qu} & $Q$ \\
\midrule
$\xi_5$         & 0.23859(11) & 0.23846(10) & 0.38 \\
$\xi_4\xi_5$ & 0.24878(14) & 0.24874(12) & 0.83 \\
$\xi_i\xi_5$ & 0.29542(69) & 0.29410(40) & 0.10 \\
$\xi_i\xi_j$ & 0.2956(11) & 0.29494(48) & 0.58 \\ 
\bottomrule
\bottomrule
\end{tabular}
\end{table}

\begin{SCfigure}
\includegraphics[width=0.5\textwidth]{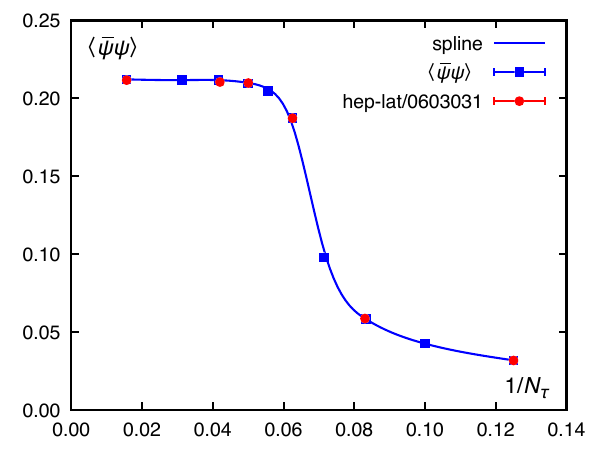}
\caption{Comparison of the chiral condensate $\langle\bar\psi\psi\rangle$ computed in Ref.~\cite{Levkova:2006gn} (red points) and in our test runs at the same parameters, summarized in Table~\ref{tab_pbp_comp} (blue points). Blue line is a spline interpolation to guide the eye.	
\label{fig_pbp_nf2_naive}}
\end{SCfigure}

Finally, to test full Rational Hybrid Monte Carlo pipeline, we generated several dynamical 2-flavor naive staggered ensembles of Ref.~\cite{Levkova:2006gn}, both at zero and finite temperature. 
The gauge action is the Wilson action and to match our definition of $\xi_0^f$, the quark mass value of Ref.~\cite{Levkova:2006gn} is multiplied by $\xi_0^f$.
On the two zero-temperature ensembles,
we measured the same pion taste, $\xi_5$ and $\xi_{4}\xi_{5}$, masses as in Ref.~\cite{Levkova:2006gn}. The comparison is shown in Table~\ref{tab_levkova_pisplit}.
The finite-temperature ensembles were generated at the same parameters as the second, $16^3\times64$ ensemble of Table~\ref{tab_levkova_pisplit}. The only varied parameter is the temporal extent $N_\tau$.
The comparison for the chiral condensate $\langle\bar\psi\psi\rangle$ is shown in Table~\ref{tab_pbp_comp} as well as in Fig.~\ref{fig_pbp_nf2_naive}. The line is not a fit but spline interpolation to guide the eye. We generated more finite-temperature ensembles to have denser temperature coverage. 
Since temperature is $T=1/(a_\tau N_\tau)$ with $a_\tau$ fixed in this case, there is no need to convert to physical units for the comparison purposes. Therefore the $x$-axis of Fig.~\ref{fig_pbp_nf2_naive} is in units of $1/N_\tau$.

\setlength{\tabcolsep}{9pt}
\begin{table}[b]
\caption{\label{tab_levkova_pisplit}
The masses of the $\xi_5$ (first line per $\beta$ value) and $\xi_0\xi_5$ (second line) pion tastes compared with the results from Ref.~\cite{Levkova:2006gn}. The first column is the gauge coupling $\beta=6/g_0^2$, second the bare gauge anisotropy, third the bare fermion anisotropy, fourth the bare quark mass, fifth the lattice volume, sixth the number of configurations used for measurements, seventh our results for the pion masses, eighth the results from Ref.~\cite{Levkova:2006gn} and ninth the $Q$ value of the Gaussian difference test.}
\begin{tabular}{l l l l l r l l l}
\toprule
\toprule
$\beta$& $\xi_0$ & $\xi_0^f$ & $a_\sigma m$ & $N_\sigma^3\times N_\tau$ & $N_{conf}$ & $a_\tau M_\pi$ & $a_\tau M_\pi$~\cite{Levkova:2006gn} & $Q$ \\
\midrule
5.425 & 1.5 & 3 & 0.0375 & $16^3\times32$ & 1,200  & 0.31303(11) & 0.31309(28) & 0.84 \\
& & & & & & 0.4687(13) & 0.4605(37) & 0.04 \\
\midrule
5.3 & 3 & 3 & 0.024 & $16^3\times64$ & 745 & 0.112641(46)  & 0.11280(51) & 0.76 \\
& & & & & & 0.14137(19) & 0.1410(11) & 0.74 \\
\bottomrule
\bottomrule
\end{tabular}
\end{table}

\setlength{\tabcolsep}{9pt}
\begin{table}
\caption{\label{tab_pbp_comp}
Comparison of the chiral condensate with the results of Ref.~\cite{Levkova:2006gn}. The parameters of the ensembles are the same as for the second ensemble in Table~\ref{tab_levkova_pisplit}, except of varying $N_\tau$, listed in the first column. The second column is the number of trajectories (with one condensate measurement per trajectory), third our values for $\langle\bar{\psi}\psi\rangle$ in lattice units, fourth values from Ref.~\cite{Levkova:2006gn} and fifth the $Q$ value of the Gaussian difference test. The values from this table are plotted in Fig.~\ref{fig_pbp_nf2_naive}.
}
\begin{tabular}{lrlll}
\toprule
\toprule
$N_\tau$ & $N_{traj}$ & $\langle\bar{\psi}\psi\rangle$ & $\langle\bar{\psi}\psi\rangle$~\cite{Levkova:2006gn} & $Q$ \\
\midrule
8 & 2,600 & 0.03191(11) & 0.03184(00) & 0.52 \\
10 & 5,090 & 0.04264(12) & - & - \\
12 & 2,600 & 0.05858(16) & 0.05876(17) & 0.44 \\
14 & 10,095 & 0.09803(55) & - & - \\
16 & 10,095 & 0.18728(43) & 0.18729(92) & 0.99 \\
18 & 9,970 & 0.20490(45) & - & - \\
20 & 5,095 & 0.20955(43) & 0.20987(43) & 0.60 \\
24 & 2,600 & 0.21174(34) & 0.21051(47) & 0.03 \\
32 & 4,795 & 0.21171(24) & - & - \\
64 & 6,980 & 0.21222(14) & 0.21171(26) & 0.08 \\
\bottomrule
\bottomrule
\end{tabular}
\end{table}

For the chiral condensate comparison we note the following. The values were not tabulated in Ref.~\cite{Levkova:2006gn} but only plotted in Fig.~9 of that reference. The Encapsulated PostScript (EPS) file of the figure is available in the arXiv version of the paper. As the EPS file is an ASCII file with a collection of drawing commands~\cite{adobe1992eps}, we can reconstruct the location of the points from it. That particular EPS file uses coordinates scaled to $[0,1]$ range with the accuracy of $0.0001$. The horizontal axis of Fig.~9 of Ref.~\cite{Levkova:2006gn} covers $[0,0.4]$ range (temperature in GeV) and the vertical $[0,0.3]$ ($\langle \bar\psi\psi\rangle$ in lattice units). This means that we can reconstruct the condensate values and errorbars from the figure with the accuracy of 0.00003. This is enough for the purposes of this paper. The other two aspects relevant for the comparison are: (a) Ref.~\cite{Levkova:2006gn} used the R algorithm, that has integration step size dependence, while we used today's standard Rational Hybrid Monte Carlo algorithm, (b) for computational efficiency the MILC code works with scaled fermion matrix $2M$ rather than $M$, so our values in Table~\ref{tab_pbp_comp} are twice the values computed by the MILC code, to match the Ref.~\cite{Levkova:2006gn} standard normalization.

\section{Testing the aHISQ fermion force}
\label{sec_app_hisq_force}
\begin{SCfigure}
\includegraphics[width=0.5\textwidth]{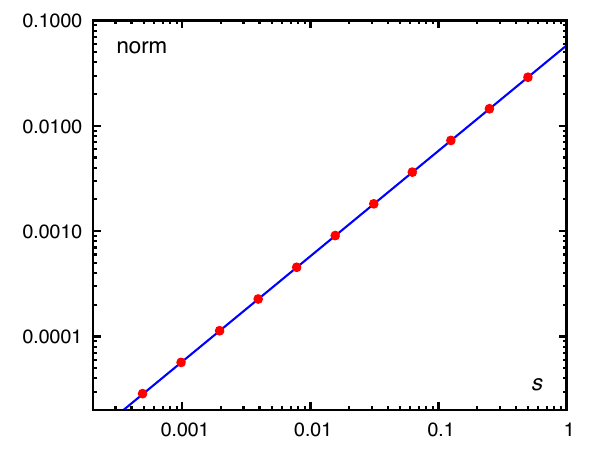}
\caption{The Frobenius norm of the difference, defined in Eq.~(\ref{eq_frobeniusmeasure}) of the aHISQ force as calculated directly in the MILC code, Eq.~(\ref{eq_fmu}), and from the finite difference (\ref{eq_force_Sf}) with $s$ ranging from $1/2$ down to $1/2048$. The measurements are on a $6^4$ lattice, at site (1,0,0,0), in the positive $\tau$ direction. The line represents a linear fit without a free term.\label{fig_force_test}}
\end{SCfigure}

Apart from the tests described in Appendix \ref{sec_app_tests}, we performed other tests of the Rational Hybrid Monte Carlo pipeline. We ran the full aHISQ RHMC, but with the smearing coefficients set to 0, to reproduce the naive staggered results. Finally, to independently test the aHISQ fermion force calculation, we compared the force evaluated by the code in the way described in Sec.~\ref{sec_dyn_ani}, where it is denoted by $f_{x,\mu}$ and defined in Eq.~(\ref{eq_fxmu_def}), with the computation of the force done by approximating the group derivative of the action with a forward finite difference:
\begin{equation}
\label{eq_force_Sf}
\tilde f_{x,\mu}=T_A \frac{d}{ds}S_f\left[e^{isT_A}U_{x,\mu}\right]\Big|_{s=0}\simeq T_A\,\frac{S_f\left[e^{isT_A}U_{x,\mu}\right]-S_f\left[U_{x,\mu}\right]}{s},
\end{equation}
for small $s$. The norm of the difference between the approximation (\ref{eq_force_Sf}) and the actual force calculation (\ref{eq_fxmu_def}) defined with the Frobenius measure as
\begin{equation}
\label{eq_frobeniusmeasure}
|| f_{x,\mu}-\tilde f_{x,\mu}||=\sqrt{\text{Tr}[(f_{x,\mu}-\tilde f_{x,\mu})^\dagger (f_{x,\mu}-\tilde f_{x,\mu})]}
\end{equation}
is shown in Fig.~\ref{fig_force_test} for $s$ ranging from $1/2$ down to $1/2048$ together with a linear fit.

\bibliography{ahisq}

\end{document}